\documentclass[aos,preprint,12pt]{imsart}

\RequirePackage[OT1]{fontenc}
\RequirePackage{amsthm,amsmath,natbib}
\RequirePackage[colorlinks]{hyperref}
\usepackage{array}
\usepackage{cleveref}
\usepackage{mathrsfs}
\usepackage{amssymb,bbm}
\usepackage{amsfonts}
\usepackage{amsmath,amssymb}
\usepackage{graphicx}
\usepackage{amscd}
\usepackage{color}
\usepackage{mathrsfs}
\usepackage{latexsym}
\usepackage{array,bm}
\usepackage{epstopdf}
\usepackage{enumerate}
\usepackage{subcaption}
\usepackage{soul}

\usepackage[flushleft]{threeparttable}

\usepackage{textcomp}

\usepackage{algorithm}
\usepackage{algorithmic}

\usepackage{rotating}
\usepackage{booktabs}
\usepackage{bigstrut, multirow}

\def\boxit#1{\vbox{\hrule\hbox{\vrule\kern6pt
          \vbox{\kern6pt#1\kern6pt}\kern6pt\vrule}\hrule}}

\newcommand{\bbeta}{\bm{\beta}}

\newcommand{\btheta}{\bm{\theta}}

\newcommand{\bI}{\mathbf{I}}

\newcommand{\bvarepsilon}{\bm{\varepsilon}}
\newcommand{\bY}{\bm{Y}}

\newcommand{\bV}{\bm{V}}

\newcommand{\bX}{\bm{X}}
\newcommand{\bS}{\bm{S}}

\newcommand{\bGamma}{\bm{\Gamma}}
\newcommand{\bpsi}{\bm{\psi}}
\newcommand{\balpha}{\bm{\alpha}}

\newcommand{\bZ}{\bm{Z}}

\newcommand{\bM}{\bm{M}}

\allowdisplaybreaks[1]

\startlocaldefs
\numberwithin{equation}{section}
\theoremstyle{plain}
\newtheorem{thm}{Theorem}[section]

\newtheorem{assum}{Assumption}[section]
\newtheorem{rem}{Remark}[section]

\endlocaldefs

\definecolor{gray}{gray}{0.5}

\begin{document}

\begin{frontmatter}
\title{Quantiled conditional variance, skewness, and kurtosis by Cornish-Fisher expansion}


\begin{aug}

 \author{\fnms{Ningning} \snm{Zhang}}
 \and
 \author{\fnms{Ke} \snm{Zhu}\thanksref{t1}
 }


\thankstext{t1}{Address correspondence to Ke Zhu: Department of Statistics and Actuarial Science, University of Hong Kong, Hong Kong.  E-mail: mazhuke@hku.hk}

 \affiliation{University of Hong Kong}

\end{aug}

%
%
\begin{abstract}
	The conditional variance, skewness, and kurtosis play a central role in time series analysis. These three conditional moments (CMs)
are often studied by some parametric models but with two big issues: the risk of model mis-specification and the instability of model estimation.
To avoid the above two issues, this paper proposes a novel method to estimate these three CMs by the so-called quantiled CMs (QCMs).
The QCM method first adopts the idea of Cornish-Fisher expansion to construct a linear regression model, based on $n$ different estimated conditional quantiles. Next, it computes the QCMs simply and simultaneously by using the ordinary least squares estimator of
this regression model, without any prior estimation of the conditional mean.
Under certain conditions, the QCMs are shown to be consistent with the convergence rate $n^{-1/2}$. Simulation studies indicate that the QCMs perform well under different scenarios of Cornish-Fisher expansion errors and quantile estimation errors.
In the application, the study of QCMs for three exchange rates demonstrates
the effectiveness of financial rescue plans during the COVID-19 pandemic outbreak, and suggests that the existing  ``news impact curve'' functions for the conditional skewness and kurtosis may not be suitable.
\end{abstract}




\end{frontmatter}
%
%
\newpage

\setcounter{equation}{0}

\section{Introduction}\label{sec_1}

Learning the conditional variance, skewness, and kurtosis of a univariate time series is the core issue
in many financial and economic applications. The classical tools to study the conditional variance are the generalized autoregressive conditional heterosecdasticity (GARCH) model and its variants. See \citet{Engle:1982}, \citet{Bollerslev:1986}, \citet{Francq:2019}, and references therein.
However, except for some theoretical works on parameter estimation in \citet{Escanciano:2009} and \citet{FT:2019}, the GARCH-type models commonly assume independent and identically distributed (i.i.d.) innovations,
resulting in the constant conditional skewness and kurtosis. As argued by
\citet{Samuelson:1970} and \citet{Rubinstein:1973}, the higher moments like skewness and kurtosis are nonnegligible, since they are not only
exemplary evidence of the non-normal returns but also relevant to the investor's optimal decision.
Along this way, a large body of literature has demonstrated the importance of conditional skewness and kurtosis
in portfolio selection (\citealp{Chunhachinda:1997}), asset pricing (\citealp{Harvey:2000}), risk management (\citealp{BMT:2008}), return predictability (\citealp{JZZ:2019}), and many others.
These empirical successes indicate the necessity to learn the dynamic structures of the conditional skewness and kurtosis simultaneously with the conditional variance.

Although there is a large number of studies on the conditional variance, only a few of them
have taken account of the conditional skewness and kurtosis. The pioneer works towards this goal are \citet{Hansen:1994} and \citet{Harvey:1999}, followed by \citet{Jondeau:2003},  \citet{BBHP:2005}, \citet{Leon:2005}, \citet{GL:2009}, and \citet{LN:2020}. All of these works assume a peculiar conditional distribution on the innovations of GARCH-type model, where
the conditional skewness or kurtosis either directly has an analogous GARCH-type dynamic structure rooting in re-scaled shocks or indirectly depends on dynamic structure of distribution parameters. See also \citet{FS:2022} and \citet{SG:2022} for a different
investigation of conditional skewness and kurtosis via the GARCH-type model with a time-varying probability of zero returns.
However,
the aforementioned parametric methods have two major shortcomings: First, they inevitably have the risk of using wrongly specified parametric models or innovation distributions; Second, they usually have unstable model estimation results in the presence of dynamic structure of skewness and kurtosis.


This paper proposes a new novel method to simultaneously learn the conditional variance, skewness, and kurtosis by the so-called quantiled
conditional variance, skewness, and kurtosis, respectively. Our three quantiled conditional moments (QCMs) are formed based on the spirit of the Cornish-Fisher expansion (\citealp{Cornish:1938}), which exhibits a fundamental relationship between the conditional quantiles and CMs. By replacing the unknown conditional quantiles with their estimators at $n$ quantile levels, the QCMs (with respect to variance, skewness, and kurtosis) are simply computed at each fixed timepoint by using the ordinary least squares (OLS) estimator of a linear regression model, which stems naturally from the Cornish-Fisher expansion. Surprisingly, our way to compute the QCMs does not require any estimator of the conditional mean.
The precision of the QCMs is controlled by the errors of the proposed linear regression model, which comprises two components: First, the expansion error encompasses higher-order conditional moments in the Cornish-Fisher expansion that are not taken as the regressors in the linear regression model; Second, the approximation error arises from the use of estimated conditional quantiles (ECQs). Under certain conditions on the regression model error,  we show that the QCMs are consistent estimators of the corresponding CMs with the convergence rate $n^{-1/2}$. Simulation studies reveal that, when considering various scenarios of approximation errors caused by the biased ECQs from the use of contaminated or mis-specified conditional quantile models, (i) the quantiled conditional variance and skewness exhibit robust and satisfactory performance,
regardless of the non-negligibility of expansion error; (ii) the quantiled conditional kurtosis has a larger dispersion for heavier-tailed data with non-negligible expansion error, which is unavoidable due to the less accuracy of CF expansion on the heavier-tailed distributions (\citealp{LeeLin:1992}).

In the application, we study the QCMs of return series for three exchange rates.
During the COVID-19 pandemic outbreak in March 2020, we find that
the values of quantiled conditional variance and kurtosis increased rapidly, and the values of quantiled conditional skewness decreased sharply
before March 19 or 20 in all of examined exchange rates, shedding light on the worldwide perilous financial crisis at that time.
After March 19 or 20, we find that the values of quantiled conditional variance, skewness, and kurtosis exhibited totally opposite trends, so demonstrating the effectiveness of financial rescue plans issued by governments.
Moreover, since the existing parametric forms of ``news impact curve'' (NIC) functions for the CMs
are chosen in an ad-hoc way (\citealp{Engle:1993}; \citealp{Harvey:1999}; \citealp{Leon:2005}),
we give a data-driven method to scrutinize the parametric forms of the NIC functions by using the QCMs. Our findings suggest that the parametric forms of NIC functions for conditional variance are appropriate, while those for conditional skewness and kurtosis may be unsuitable.




It is worth noting that our QCM method essentially transforms the problem of CM estimation to that of conditional quantile estimation. This brings us two major advantages over the aforementioned parametric methods, although we need do the conditional quantile estimation $n$ different times to implement the QCM method, and could face the risk of getting inaccurate estimate of conditional kurtosis for the very heavy-tailed data.


First, the QCM method can largely reduce the risk of model mis-specification, since the QCMs
are simultaneously computed without any prior estimation of the conditional mean, and their consistency
holds even when the specifications of conditional quantiles are mis-specified. This advantage is attractive and unexpected, since
we usually have to estimate conditional mean,
variance, skewness (or kurtosis) successively via some correctly specified parametric models.
The reason of this advantage is that the QCM method is regression-based.
Specifically, the conditional mean formally becomes one part of the intercept parameter, so it has no impact on the QCMs that are computed only from the OLS estimator of all non-intercept parameters;
meanwhile, the impact of biased ECQs from the use of wrongly specified conditional quantile models can be
aggregately offset by another part of the intercept parameter, ensuring the consistency of QCMs to a large extent. In a sense,
without specifying any parametric forms of CMs, this important feature allows us to view the QCMs as the ``observed'' CMs, and consequently, many intriguing but hardly implemented empirical studies could become tractable based on the QCMs (see, e.g., our empirical studies on NICs for the CMs).



Second, the QCM method can numerically deliver more stable estimators of the CMs than the parametric methods.
As shown in \citet{Jondeau:2003}, there exists a moment issue that places a necessary nonlinear constraint on
the conditional skewness and kurtosis, leading to a complex restriction on the admission region of model parameters.
This restriction not only raises the computational burden of parameter estimation but also makes the estimation result unstable, so it has been rarely considered in the existing parametric methods. In contrast, the QCM method directly computes the QCMs at each fixed timepoint, and this interesting feature ensures that the nonlinear constraint on the conditional skewness and kurtosis can be simply examined using the computed QCMs at each timepoint. Particularly, if this nonlinear constraint is violated at some timepoints, it is straightforward to replace the OLS estimator with a constrained least squares estimator to propose the QCMs which then satisfy the constraint automatically.

%

The remaining paper is organized as follows. Section \ref{sec_2} proposes the QCMs based on the linear regression,
and discusses the issues of conditional mean and moment constraints.
Section \ref{sec_3} establishes the asymptotics of the QCMs. Section \ref{sec_4} provides the practical implementations of the QCMs. Simulation studies are given in Section \ref{sec_5}. An application
to study the QCMs for three exchange rates and their related NICs is offered in Section \ref{sec_6}. Concluding remarks are presented in Section \ref{sec_7}. Proofs and some additional simulation results are deferred into the supplementary materials.

	\section{Quantiled Conditional Moments}\label{sec_2}
	
	\subsection{Definition}

Let $\{y_1,...,y_T\}$ be a time series of interest with length $T$, and $\mathcal{F}_t\equiv\sigma(y_s;s\leqslant t)$ be its available information set up to time $t$. Given $\mathcal{F}_{t-1}$, the conditional mean, variance, skewness, and kurtosis of $y_t$ at timepoint $t$ are defined as $\mu_t = E(y_t|\mathcal{F}_{t-1})$ and
	\begin{equation}
		\centering
		\begin{aligned}
			h_t = E[(y_t-\mu_t)^2|\mathcal{F}_{t-1}],\text{ }s_t = E\Big(\Big(\frac{y_t-\mu_t}{\sqrt{h_t}}\Big)^3|\mathcal{F}_{t-1}\Big),\text{ } k_t = E\Big(\Big(\frac{y_t-\mu_t}{\sqrt{h_t}}\Big)^4|\mathcal{F}_{t-1}\Big),
		\end{aligned}
		\label{cm}
	\end{equation}
respectively. Below, we show how to estimate these three conditional moments in (\ref{cm}) by using the
 Cornish-Fisher expansion (\citealp{Cornish:1938}) at a fixed timepoint $t$.

Let $Q_t(\alpha)$ be the conditional quantile of $y_t$ at the quantile level $\alpha\in(0,1)$. According to the Cornish-Fisher expansion, we have
	\begin{equation}\label{cf}
		Q_t(\alpha) = \mu_t+\sqrt{h_t}\Big[x + (x^2-1)\frac{s_t}{6} + (x^3-3x)\frac{k_t-3}{24}+r_t(\alpha)\Big],
	\end{equation}
where $x=\Phi^{-1}(\alpha)$ with $\Phi(\cdot)$ being the distribution function of $N(0, 1)$, and $r_t(\alpha)$ contains all remaining terms on the higher-order conditional moments. Taking $n$ quantile levels $\alpha_i$, $i = 1, ..., n$, equation (\ref{cf}) entails the following regression model with deterministic explanatory variables $\bX_i$ but random coefficients $\bbeta_t$:
\begin{align}\label{reg1}
Y_{t,i}^{\ast}=\mu_t+\bX_i'\bbeta_t+\varepsilon_{t,i}^{\ast}, \,\,i=1,...,n,
\end{align}
where $Y_{t,i}^{\ast}=Q_{t}(\alpha_i)$, $\bX_i=(x_i, x_i^2-1, x_i^3-3x_i)'$ with $x_i=\Phi^{-1}(\alpha_i)$,
\begin{align}\label{coeffcients}
\bbeta_t\equiv(\beta_{1t}, \beta_{2t}, \beta_{3t})'=\Big(\sqrt{h_t}, \frac{\sqrt{h_t}s_t}{6}, \frac{\sqrt{h_t}(k_t-3)}{24}\Big)',
\end{align}
and $\varepsilon_{t,i}^{\ast}=\sqrt{h_t}r_t(\alpha_i)$.
We call $\varepsilon_{t,i}^{\ast}$ the expansion error, since it comes from the Cornish-Fisher expansion but can not be explained by $\bX_i$ adequately.
%

Next, we aim to obtain the estimators of $h_t$, $s_t$, and $k_t$ through the estimator of $\bbeta_t$ in (\ref{coeffcients}). To achieve this goal, we replace the unobserved $Y_{t,i}^{\ast}$ with its estimator $Y_{t,i}$, and then rewrite model (\ref{reg1}) as follows:
\begin{align}\label{reg2}
Y_{t,i}=\mu_t+\bX_i'\bbeta_t+\varepsilon_{t,i}^{\bullet}, \,\,i=1,...,n,
\end{align}
where $Y_{t,i}=\widehat{Q}_{t}(\alpha_i)$ with $\widehat{Q}_{t}(\alpha_i)$ being an estimator of $Q_{t}(\alpha_i)$, and
$\varepsilon_{t,i}^{\bullet}=\varepsilon_{t,i}^{\ast}+\varepsilon_{t,i}^{\circ}$ with $\varepsilon_{t,i}^{\circ}=\widehat{Q}_{t}(\alpha_i)-Q_{t}(\alpha_i)$.
Clearly, $\varepsilon_{t,i}^{\circ}$ quantifies the error caused by using $\widehat{Q}_{t}(\alpha_i)$ to approximate $Q_{t}(\alpha_i)$, so it can be termed as the approximation error. Consequently, $\varepsilon_{t,i}^{\bullet}$ considering the total number of
$\varepsilon_{t,i}^{\ast}$ and $\varepsilon_{t,i}^{\circ}$ can be viewed as the gross error.
We should mention that
any two quantile levels $\alpha_i$ and $\alpha_j$ in (\ref{reg2}) are allowed to be the same, as long as $Y_{t,i}$ and $Y_{t,j}$ are different due to the use of two different conditional quantile estimation methods.
In other words, model (\ref{reg2}) allows us to simply pool different information of conditional quantiles from different channels of estimation methods at any fixed quantile level.

Although $\varepsilon_{t,i}^{\bullet}$ is expected to have values oscillating around zero, it may not always have mean zero.
Therefore, for the purpose of identification, we add a deterministic term $\gamma_t$ into
model (\ref{reg2}) to form the following regression model:
\begin{align}\label{reg3}
Y_{t,i}&=(\mu_t+\gamma_t)+\bX_i'\bbeta_t+\varepsilon_{t,i} \equiv \bZ_i'\btheta_t+\varepsilon_{t,i},\,\,i=1,...,n,
\end{align}
where $\varepsilon_{t,i}=\varepsilon_{t,i}^{\bullet}-\gamma_t$, $\bZ_i=(1,\bX_i')'$, and $\btheta_t=(\beta_{0t},\bbeta_t')'$ with
the intercept parameter $\beta_{0t}=\mu_t+\gamma_t$.

Let $\bY_t$ be an $n\times 1$ vector with entries $Y_{t,i}$, $\bZ$ be an $n\times 4$ matrix with rows $\bZ_i'$, and $\bvarepsilon_t$ be an $n\times 1$ vector with entries $\varepsilon_{t,i}$. Then, the ordinary least squares (OLS) estimator of
$\btheta_t$ in (\ref{reg3}) is
\begin{align}\label{ols}
\widehat{\btheta}_t\equiv(\widehat{\beta}_{0t},\widehat{\bbeta}_t')'=(\bZ'\bZ)^{-1}\bZ'\bY_t.
\end{align}
According to (\ref{coeffcients}), we rationally use $\widehat{\bbeta}_t\equiv (\widehat{\beta}_{1t}, \widehat{\beta}_{2t}, \widehat{\beta}_{3t})'$ in (\ref{ols}) to propose the estimators $\widehat{h}_t$, $\widehat{s}_t$, and $\widehat{k}_t$ for $h_t$, $s_t$, and $k_t$, respectively, where
\begin{equation}\label{qcm}
		\widehat{h}_t = \widehat{\beta}_{1t}^{\,2},\,\,\widehat{s}_t = \frac{6\widehat{\beta}_{2t}}{\widehat{\beta}_{1t}},\text{ and }\widehat{k}_t=\frac{24\widehat{\beta}_{3t}}{\widehat{\beta}_{1t}}+3.
	\end{equation}
We call $\widehat{h}_t$, $\widehat{s}_t$, and $\widehat{k}_t$ the quantiled conditional variance, skewness, and kurtosis of $y_t$, since they are estimators of $h_t$, $s_t$, and $k_t$, by using the estimated conditional quantiles (ECQs) of $y_t$.
Clearly, provided  $n$ different ECQs (that is, $n$ different $Y_{t,i}$ in (\ref{reg3})), our three quantiled conditional moments (QCMs) in (\ref{qcm}) are easy-to-implement, since their computation only relies on the OLS estimator $\widehat{\btheta}_t$.

\subsection{The conditional mean issue}\label{sec_2_2}

Using $\widehat{\beta}_{0t}$ in (\ref{ols}), we can estimate $\beta_{0t}$ but not $\mu_t$ due to the presence of $\gamma_t$.
Hence, we are unable to form the quantiled conditional mean of $y_t$ to estimate $\mu_t$.
Interestingly, our way to compute $\widehat{h}_t$, $\widehat{s}_t$, and $\widehat{k}_t$ does not require an estimator of $\mu_t$. This is
far beyond our expectations, since normally we have to first estimate (or model) $\mu_t$ and then $h_t$, $s_t$, and $k_t$.

Although the estimation of $\mu_t$ is not required to compute $\widehat{h}_t$, $\widehat{s}_t$, and $\widehat{k}_t$, knowing the dynamic structure of $\mu_t$ is still important in practice. Note that $\mu_t$ is often assumed to be an unknown constant in accordance with the efficient market hypothesis, and this constant assumption can be examined by the consistent spectral tests for the martingale difference hypothesis (MDH) in \citet{EV:2006}. If the constant assumption is rejected by these tests, the dynamic structure of $\mu_t$ manifests and is usually specified by
a linear model (e.g., the autoregressive moving-average model) or a nonlinear model (e.g., the threshold autoregressive model);
see \citet{FY:2003} and \citet{Tsay:2005} for surveys.
In this case, the model correctly specifies the dynamic structure of $\mu_t$ if and only if
its model error is an MD sequence, a statement which can be consistently checked by two spectral tests
for the MDH on unobserved model errors in \citet{Escanciano:2006}.
Hence, it is usually tractable for practitioners to come up with a valid parametric model for $\mu_t$ in most of applications.


\subsection{The moment constraints issue}

Note that $\mu_t$, $h_t$, $s_t$, and $k_t$ can be expressed in terms of the first four non-central moments $m_{1t}$, $m_{2t}$, $m_{3t}$, and $m_{4t}$ of $Y_t$, where $m_{jt}=E(Y_t^{j}|\mathcal{F}_{t-1})$. Therefore, the existence of $\mu_t$, $h_t$, $s_t$, and $k_t$ is equivalent to that of
$m_{1t}$, $m_{2t}$, $m_{3t}$, and $m_{4t}$, and the latter requires the existence of a non-decreasing function $F_t(\cdot)$ such that
$$m_{jt}=\int_{-\infty}^{\infty}x^j dF_t(x).$$
To ensure this existence, Theorem 12.a in \citet{Widder:1946} indicates that the following condition must hold for $m_{1t}$, $m_{2t}$, $m_{3t}$, and $m_{4t}$:
\begin{equation}\label{check_cond}
\det\begin{pmatrix}
m_{0t} & m_{1t}\\
m_{1t} & m_{2t}
\end{pmatrix}\geq 0
\,\,\,\mbox{ and }
\,\,\,
\det\begin{pmatrix}
m_{0t} & m_{1t} & m_{2t}\\
m_{1t} & m_{2t} & m_{3t}\\
m_{2t} & m_{3t} & m_{4t}
\end{pmatrix}\geq 0.
\end{equation}
By some direct calculations, it is not hard to see that condition (\ref{check_cond}) is equivalent to
\begin{equation}\label{equiv_cond}
h_t\geq 0\,\,\,\mbox{ and } \,\,\, k_t-s_t^2-1\geq 0.
\end{equation}
Condition (\ref{equiv_cond}) above places two necessary moment constraints on $h_t$, $s_t$, and $k_t$.
When $h_t$, $s_t$, and $k_t$ are specified by some parametric models with unknown parameters,
the first moment constraint usually can be easily handled, but
the second moment constraint restricts the admission region of unknown parameters in a very complex way, so that the model estimation becomes quite inconvenient and unstable. This is the reason why the second moment constraint has been rarely taken into account in the literature except \citet{Jondeau:2003}.

Impressively, the moment constraints issue above is not an obstacle for our QCMs, since the QCMs estimate the CMs directly at each fixed timepoint $t$. In view of the relationship between the QCMs and $\widehat{\bbeta}_t$ in (\ref{qcm}), we know that the QCMs satisfy two constraints in (\ref{equiv_cond}) if and only if $\widehat{\beta}_{1t}^2\geq 0$ and $\widehat{\beta}_{1t}^2-18\widehat{\beta}_{2t}^2+12\widehat{\beta}_{1t}\widehat{\beta}_{3t}\geq 0$.
Since the constraint $\widehat{\beta}_{1t}^2\geq 0$ holds automatically, we indeed only need to check whether
\begin{equation}\label{equiv_cond_qcr}
\widehat{\beta}_{1t}^2-18\widehat{\beta}_{2t}^2+12\widehat{\beta}_{1t}\widehat{\beta}_{3t}\geq 0.
\end{equation}
In practice, the constraint in (\ref{equiv_cond_qcr}) can be directly examined after the QCMs are computed. Our applications in Section \ref{sec_6} below show that this constraint holds at all examined timepoints.
In other applications, if the constraint in (\ref{equiv_cond_qcr}) does not hold at some timepoints $t$, we can easily re-estimate $\btheta_t$ in (\ref{reg3}) by the constrained least squares estimation method with the constraint
$\beta_{1t}^2-18\beta_{2t}^2+12\beta_{1t}\beta_{3t}\geq 0$, so that the resulting QCMs satisfy the constraint in (\ref{equiv_cond_qcr}) automatically at these timepoints.

\section{Asymptotics}\label{sec_3}

This section studies the asymptotics of $\widehat{h}_t$, $\widehat{s}_t$, and $\widehat{k}_t$ at a fixed timepoint $t$.
Let $\overset{p}{\longrightarrow}$ denote convergence in probability.
To derive the consistency of $\widehat{h}_t$, $\widehat{s}_t$, and $\widehat{k}_t$ in (\ref{qcm}), the following assumptions are needed.

\begin{assum}\label{cond1}
$\bM_n\equiv\bZ'\bZ/n$ is uniformly positive definite.
\end{assum}

\begin{assum}\label{cond2}
$\bZ'\bvarepsilon_{t}/n\overset{p}{\longrightarrow} \pmb{0}$ as $n\to\infty$.
\end{assum}

We offer some remarks on the aforementioned assumptions. Assumption \ref{cond1} is regular for linear regression models,
and it holds as long as $\bZ$ is fully ranked (i.e., Rank($\bZ$)=4).
Because $\varepsilon_{t,i}=\varepsilon_{t,i}^{\ast}+\varepsilon_{t,i}^{\circ}-\gamma_{t}$, Assumption \ref{cond2} is equivalent to
\begin{align}
&C_{t,\ast}+C_{t,\circ}-C_{t,\gamma}\overset{p}{\longrightarrow} \pmb{0} \mbox{ as }n\to\infty, \label{cond_circ}
\end{align}
where $C_{t,\ast}=n^{-1}\sum_{i=1}^{n}\bZ_i\varepsilon_{t,i}^{\ast}$,  $C_{t,\circ}=n^{-1}\sum_{i=1}^{n}\bZ_i\varepsilon_{t,i}^{\circ}$, and $C_{t,\gamma}=\big(n^{-1}\sum_{i=1}^{n}\bZ_i\big)\gamma_{t}$.
By law of large numbers for dependent and heteroscedastic data sequence (\citealp{Andrews:1988}), it is reasonable to assert that $C_{t,\ast}=n^{-1}\sum_{i=1}^{n}\bZ_iE(\varepsilon_{t,i}^{\ast})+o_{p}(1)$ and $C_{t,\circ}=n^{-1}\sum_{i=1}^{n}\bZ_iE(\varepsilon_{t,i}^{\circ})+o_{p}(1)$. Then,
since $\varepsilon_{t,i}^{\bullet}=\varepsilon_{t,i}^{\ast}+\varepsilon_{t,i}^{\circ}$,
condition (\ref{cond_circ}) holds if
\begin{align}\label{eqn_gamma}
\frac{1}{n}\sum_{i=1}^{n}\bZ_i\big[E(\varepsilon_{t,i}^{\bullet})-\gamma_{t}\big]\longrightarrow \pmb{0} \mbox{ as }n\to\infty.
\end{align}
Condition (\ref{eqn_gamma}) reveals an important fact that the role of $\gamma_{t}$ is to offset the possible
non-identification effect caused by the non-zero mean of $\varepsilon_{t,i}^{\bullet}$.
In other words,  to achieve the identification, $\gamma_{t}$ should automatically tend to minimize the absolute difference
$$d_{n,t}\equiv\Big|\frac{1}{n}\sum_{i=1}^{n}\bZ_i\big[E(\varepsilon_{t,i}^{\bullet})\big]-\Big(\frac{1}{n}\sum_{i=1}^{n}\bZ_i\Big)\gamma_{t}\Big|$$
for large $n$. Clearly, if $d_{n,t}\approx 0$ for large $n$, Condition (\ref{eqn_gamma}) holds automatically, so then Assumption \ref{cond2} most likely holds.

Next, we study the behavior of $d_{n,t}$ in different cases. For the first case that
$E(\varepsilon_{t,i}^{\bullet})\approx c_t$ for all $i$, we have $d_{n,t}\approx 0$ with $\gamma_t= c_t$.
For the second case that $E(\varepsilon_{t,i}^{\bullet})\approx 0$ for most of $i$, we also have $d_{n,t}\approx 0$ with $\gamma_t=0$.
For the third case that $n^{-1}\sum_{i=1}^{n}\bZ_i\big[E(\varepsilon_{t,i}^{\bullet})\big]\approx\pmb{\tau}_t\approx\pmb{z} f_t$
and $n^{-1}\sum_{i=1}^{n}\bZ_i\approx\pmb{z}$ for large $n$, we again have $d_{n,t}\approx 0$ with $\gamma_t=f_t$.
For other cases, we still have the chance to ensure $d_{n,t}\approx 0$, depending on the
behavior of  $E(\varepsilon_{t,i}^{\bullet})$ across $i$. In summary, the condition that
 $E(\varepsilon_{t,i}^{\bullet})\approx 0$ for all $i$ is not necessary for the validity of Assumption \ref{cond2}. This implies that the QCMs are able to have robust performances across diverse error scenarios, including situations where $E(\varepsilon_{t,i}^{\circ})$ has the large non-zero absolute values across $i$ (that is, the large biases of ECQs caused by the use of mis-specified conditional quantile models).

As expected, the condition that $d_{n,t}\approx 0$ for large $n$ may not always hold, and therefore, under certain circumstances, Assumption \ref{cond2} may fail. For example, when the tail of $y_t$ becomes heavier, the impact of higher-order conditional moments in the Cornish-Fisher expansion becomes larger. In this case, $E(\varepsilon_{t,i}^{*})$ tends to have a more non-negligible exotic behavior across $i$, so that it is harder to offset the non-identification effect via $\gamma_t$, with the value of $d_{n,t}$ farther away from zero. Indeed, our simulation studies in Section \ref{sec_5} below indicate that the presence of non-negligible $\varepsilon_{t,i}^{*}$ has a larger impact on the consistency of $\widehat{k}_t$ than $\widehat{h}_t$ and $\widehat{s}_t$,  which perform robustly in terms of the heavy-tailedness of $y_t$.

The following theorem establishes the consistency of $\widehat{h}_t$, $\widehat{s}_t$, and $\widehat{k}_t$.

\begin{thm}\label{thm1}
Suppose that Assumptions \ref{cond1}--\ref{cond2} hold. Then, $\widehat{\btheta}_{t}-\btheta_{t}\overset{p}{\longrightarrow} 0$ as $n\to\infty$. Consequently,
$\widehat{h}_t-h_t\overset{p}{\longrightarrow} 0$, $\widehat{s}_t-s_t\overset{p}{\longrightarrow} 0$, and $\widehat{k}_t-k_t\overset{p}{\longrightarrow} 0$ as $n\to\infty$.
\end{thm}

\begin{rem}
If we assume $\bZ'\bvarepsilon_{t}/n\overset{}{\longrightarrow} \pmb{0}$ almost surely as $n\to\infty$ in Assumption \ref{cond2}, all of convergence results in Theorem \ref{thm1} hold almost surely.
\end{rem}

\begin{rem}
In Theorem \ref{thm1}, we require large $n$ but not large $T$. Certainly, a large $T$
may improve the performance of ECQs with small biases, however, it is not necessary for the validity of Assumption \ref{cond2} and thus the
consistency of QCMs.
\end{rem}

Let $\overset{d}{\longrightarrow}$ denote convergence in distribution. We raise the following higher order assumption to replace Assumption \ref{cond2}:

\begin{assum}\label{cond3}
$[\bV_{t,n}]^{-1/2}\big[\bZ'\bvarepsilon_{t}/\sqrt{n}\big]\overset{d}{\longrightarrow} N(0, \bI)$ as $n\to\infty$, where $\bI$ is an identity matrix, and $\bV_{t,n}\equiv var(\bZ'\bvarepsilon_{t}/\sqrt{n})$ is bounded and uniformly positive definite.
\end{assum}

Assumption \ref{cond3} is regular for proving the asymptotic normality of OLS estimator (see \citet{White:2001}).
The theorem below shows that $\widehat{h}_t$, $\widehat{s}_t$, and $\widehat{k}_t$ are $\sqrt{n}$-consistent but not asymptotically normal.

\begin{thm}\label{thm2}
Suppose that Assumptions \ref{cond1} and \ref{cond3} hold. Then,
$$(\bM_n^{-1}\bV_{t,n}\bM_n^{-1})^{-1/2}\sqrt{n}(\widehat{\btheta}_{t}-\btheta_{t})\overset{d}{\longrightarrow} N(0, \bI)$$
as $n\to\infty$. Moreover,
$\sqrt{n}(\widehat{h}_t-h_t)=O_p(1)$, $\sqrt{n}(\widehat{s}_t-s_t)=O_p(1)$, and $\sqrt{n}(\widehat{k}_t-k_t)=O_p(1)$, but
$\widehat{h}_t$, $\widehat{s}_t$, and $\widehat{k}_t$ are not asymptotically normal.
\end{thm}

\begin{rem}
Although $\widehat{h}_t$, $\widehat{s}_t$, and $\widehat{k}_t$ are not asymptotically normal, the asymptotic normality of  $\widehat{\btheta}_{t}$ demonstrates that the quantiled volatility (the second entry of $\widehat{\btheta}_{t}$), denoted by $\widehat{\sigma}_t$, has the asymptotic normality property:
$(\bGamma\bM_n^{-1}\bV_{t,n}\bM_n^{-1}\bGamma')^{-1/2}\sqrt{n}(\widehat{\sigma}_t-\sigma_t)\overset{d}{\longrightarrow} N(0, 1)$ as $n\to\infty$,
where $\bGamma=(0, 1, 0, 0)$.
\end{rem}

As shown before, the asymptotics of QCMs in Theorems \ref{thm1}--\ref{thm2} hold with no need to consider the specification of conditional mean and choose the specifications of conditional quantiles correctly. This important feature guarantees that the QCM method can largely reduce the risk of model mis-specification. The reason of this feature is that the QCM method is regression-based. Specifically, the conditional mean $\mu_{t}$ is absorbed into the intercept parameter $\beta_{0,t}$, so that it has no impact on the QCMs; meanwhile, the biases of ECQs
from the use of wrongly specified conditional quantile models can be aggregately offset by the term $\gamma_t$, which is also nested
in $\beta_{0,t}$.

The aforementioned feature is accompanied by two limitations. The first limitation is that we need to estimate conditional quantiles $n$ different times. Fortunately, this limitation seems mild, since the quantile estimation commonly can be computed easily by the linear programming method and the resulting estimation biases can be tolerated by the QCM method to a large extent. The second limitation of the QCMs is that when the data are very heavy-tailed, the expansion error could have a non-negligible impact to cause the identification problem, particularly for the conditional kurtosis. This limitation seems an unavoidable consequence of the Cornish-Fisher expansion, and it could not be addressed by simply increasing the order of expansion (\citealp{LeeLin:1992}).

\section{Practical Implementations}\label{sec_4}
To compute three QCMs in (\ref{qcm}), we only need to input $n$ different ECQs $\widehat{Q}_{t}(\alpha_i)$, which can be computed in many different ways; see, for example, \cite{MF:2000}, \cite{KMP:2006}, \cite{XK:2009} and the references therein for earlier works, and \cite{KCHP:2017} and \cite{Zheng:2018} for more recent ones. Without assuming any parametric specifications of CMs, \citet{Engle:2004} propose a general class of CAViaR models, which can decently specify the dynamic specification of conditional quantiles. Hence, the CAViaR models are appropriate choices for us to compute $\widehat{Q}_{t}(\alpha_i)$. Following \citet{Engle:2004}, we consider four CAViaR models below:


\begin{enumerate}
\item [(1)] \textit{Symmetric Absolute Value (SAV) model:} $Q_t(\alpha) = \psi_{1,0} + \psi_{2,0}Q_{t-1}(\alpha)+\psi_{3,0}|y_{t-1}|$;

\item [(2)] \textit{Asymmetric Slope (AS) model:} $Q_t(\alpha) = \psi_{1,0} + \psi_{2,0}Q_{t-1}(\alpha)+\psi_{3,0}(y_{t-1})^{+}+\psi_{4,0}(y_{t-1})^{-}$, where $(y_{t-1})^{+}=\text{max}(y_{t-1},0)$ and $(y_{t-1})^{-}=\text{min}(y_{t-1},0)$;
\item [(3)]	\textit{Indirect GARCH (IG) model:} $Q_t(\alpha) = (\psi_{1,0} + \psi_{2,0}Q^2_{t-1}(\alpha)+\psi_{3,0}y^2_{t-1})^{1/2}$;

\item [(4)]	\textit{Adaptive (ADAP) model:} $Q_t(\alpha) = Q_{t-1}(\alpha) + \psi_{1,0}\{[1+\text{exp}(N[y_{t-1}-Q_{t-1}(\alpha)])]^{-1}-\alpha\}$, where $N$ is a positive finite number.
\end{enumerate}

Each CAViaR model above can be estimated via the classical quantile estimation method (\citealp{Koenker:1978}).
For simplicity, we take the SAV model as an illustrating example.
Let $\bpsi=(\psi_1, \psi_{2},\psi_{3})'$ be the unknown parameter of SAV model, and
$\bpsi_0=(\psi_{1,0}, \psi_{2,0}, \psi_{3,0})'$ be its true value. As in \citet{Engle:2004}, we estimate $\bpsi_0$ by the quantile estimator
$\widehat{\bpsi}_n\equiv\arg\min_{\bpsi} \rho_{\alpha}(y_t-Q_t(\alpha,\bpsi))$,
where $\rho_{\alpha}(x)=x[\alpha-\mbox{I}(x<0)]$ is the check function,
and $Q_t(\alpha,\bpsi)$ is defined in the same way as $Q_t(\alpha)$ in the SAV model with $\bpsi_0$ replaced by
$\bpsi$. Once $\widehat{\bpsi}_n$ is obtained, we then take $\widehat{Q}_{t}(\alpha)\equiv Q_{t}(\alpha,\widehat{\bpsi}_n)$ as
our ECQ.

Using the above CAViaR models, we can obtain different $\widehat{Q}_t(\alpha)$. However, at some quantile levels $\alpha$, some of models may be inadequate to specify the dynamic structure of $Q_t(\alpha)$, resulting in invalid $\widehat{Q}_t(\alpha)$. To screen out those invalid $\widehat{Q}_t(\alpha)$ before computing the QCMs, we consider the in-sample dynamic quantile (DQ) test $\mbox{DQ}_{IS}(\alpha)$ in Section 6 of \citet{Engle:2004}. This test $\mbox{DQ}_{IS}(\alpha)$ aims to
 detect the inadequacy of CAViaR models by examining whether
 $\bar{\bX}_t(\alpha)\mbox{Hit}_t(\alpha)$ has mean zero, where  $\mbox{Hit}_t(\alpha)\equiv \mbox{I}(y_t<Q_t(\alpha))-\alpha$ and
 $\bar{\bX}_t(\alpha)\equiv (\mbox{Hit}_{t-1}(\alpha),...,\mbox{Hit}_{t-4}(\alpha))'$. The testing idea of  $\mbox{DQ}_{IS}(\alpha)$ relies on the fact that $\bar{\bX}_t(\alpha)\mbox{Hit}_t(\alpha)$ has mean zero when the CAViaR model specifies
the dynamic structure of $Q_t(\alpha)$ correctly. Based on the sequence
 $\{\widehat{Q}_{t}(\alpha)\}_{t=1}^{T}$ from a given CAViaR model, we can compute $\mbox{DQ}_{IS}(\alpha)$ and then its
  p-value $P(\xi>\mbox{DQ}_{IS}(\alpha))$, where $\xi\sim \chi_4^2$ (a Chi-squared distribution with $4$ degrees of freedom).
 If the p-value of $\mbox{DQ}_{IS}(\alpha)$ is less than $p^*$, the corresponding ECQs
 $\{\widehat{Q}_{t}(\alpha)\}$ are deemed to be invalid, so they are not included to compute the QCMs.

Below, we summarize our aforementioned procedure to compute the QCMs:

\begin{description}
  \item[Procedure 4.1. (The steps to compute $\widehat{h}_t$, $\widehat{s}_t$, and $\widehat{k}_t$)]\
\begin{enumerate}
\item Obtain $\{\widehat{Q}_{t}(\alpha)\}_{t=1}^{T}$ at any quantile level $\alpha$ in $\balpha$ based on any CAViaR model in $\mathcal{\bM}$, where $\balpha\equiv[0.01:0.01:0.99]$ is a sequence of real numbers from $0.01$ to $0.99$ incrementing by $0.01$, and
$\mathcal{\bM}\equiv\{\mbox{SAV},\,\, \mbox{AS},\,\, \mbox{IG},\,\, \mbox{ADAP}\}$.

\item Apply the DQ test $\mbox{DQ}_{IS}(\alpha)$ to each $\{\widehat{Q}_{t}(\alpha)\}_{t=1}^{T}$ from Step 1, and discard those
$\{\widehat{Q}_{t}(\alpha)\}_{t=1}^{T}$ with p-values of $\mbox{DQ}_{IS}(\alpha)$ less than $p^*$.

\item Group all remaining $\widehat{Q}_{t}(\alpha)$ to form a set $\bS_t$ at each given $t$. Then, take the
$i$-th entry of $\bS_t$ to be $Y_{t,i}$ in (\ref{reg3}), and use its corresponding quantile level to compute $\bX_i$ in (\ref{reg3}), where
$i=1,...,n_0$, and $n_0$ is the size of $\bS_t$.

\item Based on $\{Y_{t,i}, \bX_i\}_{i=1}^{n_0}$ from Step 3, compute the OLS estimator $\widehat{\btheta}_{t}$ in (\ref{ols}) and then the three QCMs $\widehat{h}_t$, $\widehat{s}_t$, and $\widehat{k}_t$  in (\ref{qcm}).
\end{enumerate}
\end{description}

In Procedure 4.1, the value of $n_0$ decreases with that of $p^*$, and it achieves the upper bound $n=99\times 4$ when $p^*=0$ (i.e., no ECQs are discarded).
Clearly, the choice of $p^*$ reveals a trade-off between estimation reliability and estimation efficiency in the QCM method, since
a large value of $p^*$ enhances the reliability of $\widehat{Q}_{t}(\alpha)$, but meanwhile,
it reduces the efficiency of $\widehat{\btheta}_{t}$ as the value of $n_0$ becomes small.
So far, how to choose $p^*$ optimally is unclear.
Our additional simulation results in the supplementary materials show that $p^*=0.1$ is a good choice to balance the bias and variance of the estimation error of the QCMs. Hence, we recommend to take $p^*=0.1$ for the practical use.

\section{Simulations}\label{sec_5}


This section examines the finite sample performance of three QCMs $\widehat{h}_t$, $\widehat{s}_t$, and $\widehat{k}_t$.
 For saving the space, some additional simulation results
on the selection of $p^*$ are reported in the supplementary materials.

\subsection{Simulations on the GARCH model}
We examine the performance of QCMs when the data are generated by the benchmark GARCH model.
Specifically, we generate 100 replications of sample size $T=1000$ from the following GARCH model (\citealp{Bollerslev:1986})
\begin{equation}
\label{simgarch}
y_t=\eta_t\sigma_t \mbox{ and }\sigma_t^2=\omega+\alpha y_{t-1}^2+\beta\sigma_{t-1}^2,
\end{equation}
where the parameters are chosen as $\omega=0.1$, $\alpha=0.1$, and $\beta=0.8$,
and $\{\eta_t\}_{t=1}^{T}$ is an i.i.d. sequence with $\eta_t\sim N(0, 1)$ and $ST_{\nu_t}$. Here, $ST_{\nu}$ is the standardized $t_{\nu}$ distribution with mean zero and variance one, and $\nu_t$ is generated from the Uniform distribution over the interval $[5,20]$. For each replication, we can easily see that when $\eta_t\sim N(0, 1)$, the true values of CMs and conditional quantiles of $y_t$ in (\ref{simgarch}) are
\begin{equation*}
\mu_t=0, h_t=\sigma_t^2, s_t=0, k_t=3, Q_{t}(\alpha)=\sigma_t\times\mbox{quantile of }N(0,1)\mbox{ at level }\alpha;
\end{equation*}
when $\eta_t\sim ST_{\nu_t}$, those of $y_t$ are
\begin{equation*}
\mu_t=0, h_t=\sigma_t^2, s_t=0, k_t=\frac{6}{\nu_t-4}+3, Q_{t}(\alpha)=\sigma_t\times\mbox{quantile of }ST_{\nu_t}\mbox{ at level }\alpha.
\end{equation*}
Note that the expansion errors $\varepsilon_{t,i}^{*}$ from the Cornish-Fisher expansion in regression model (\ref{reg3}) are negligible when the conditional distribution of $y_t$ is normal in the case of $\eta_t\sim N(0, 1)$, whereas $\varepsilon_{t,i}^{*}$ are non-negligible when the conditional distribution of $y_t$ is
heavy-tailed in the case of $\eta_t\sim ST_{\nu_t}$; see \citet{LeeLin:1992}.

Next, we generate the sequence $\{\widehat{Q}_{t}(\alpha_i)\}$ at each $t$ to compute $\widehat{h}_t$, $\widehat{s}_t$, and $\widehat{k}_t$ in four different cases:
\begin{equation}  \label{error_cases}
		\begin{aligned}
			&\mbox{Case 1 [No Error]:}  \,\,\widehat{Q}_{t}(\alpha_i)=Q_{t}(\alpha_i); \\
             &\mbox{Case 2 [Error I]:}  \,\,\,\,\,\,\,\widehat{Q}_{t}(\alpha_i)=Q_{t}(\alpha_i)+\varepsilon_{t,i}^{\circ} \mbox{ with }\varepsilon_{t,i}^{\circ}\sim N(0,\sigma^2(\alpha_i));\\
			&\mbox{Case 3 [Error II]:}  \,\,\,\,\,\widehat{Q}_{t}(\alpha_i)=Q_{t}(\alpha_i)+\varepsilon_{t,i}^{\circ} \mbox{ with }\varepsilon_{t,i}^{\circ}\sim N(\mu(\alpha_i),\sigma^2(\alpha_i));\\
&\mbox{Case 4 [CAViaR]:}  \,\,\,\,\widehat{Q}_{t}(\alpha_i) \mbox{ is the entry of } \bS_t,
		\end{aligned}
	\end{equation}
where $\alpha_i\in\balpha$ in Cases 1--3, $\balpha$ and $\bS_t$ are defined as in Procedure 4.1,
 $Q_t(\alpha)$ is the true value of conditional quantile of $y_t$ at level $\alpha$, $\sigma^2(\alpha)=0.5\sigma^2+|\alpha-0.5|\sigma^2$ with $\sigma^2=0.2$, and
$\mu(\alpha)=\exp(-200\alpha)\mbox{I}(\alpha<0.5)+\exp(-(2-200\alpha))\mbox{I}(\alpha\geq0.5)$. Under the setting in Case 1, there has no approximation errors $\varepsilon_{t,i}^{\circ}$ in regression model (\ref{reg3}). Under the settings in Cases 2 and 3, the approximation errors $\varepsilon_{t,i}^{\circ}$ are considered
with different variances across $\alpha_i$, where their means are zeros (Case 2) or non-zero values (Case 3) to mimic the scenario that
$\widehat{Q}_{t}(\alpha_i)$ is an unbiased or biased estimator of $Q_{t}(\alpha_i)$, respectively.
Under the setting in Case 4, $\widehat{Q}_{t}(\alpha_i)$ are computed by the CAViaR models, and this case mimics the
real application scenario that the dynamic structures of $Q_{t}(\alpha_i)$ are unknown and modelled by the CAViaR models.

\begin{figure}[!h]
\centering
\includegraphics[width=150mm,height=120mm]{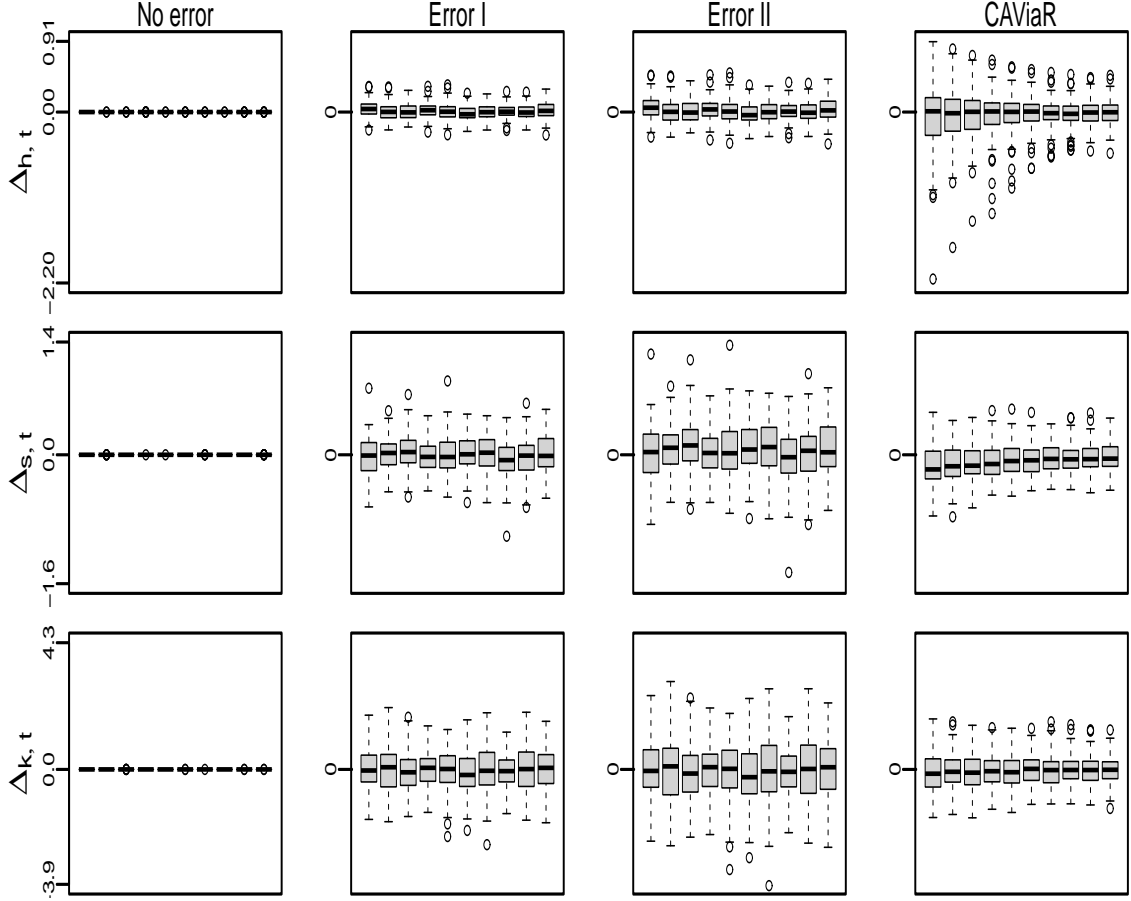}
\caption{The boxplots of $\Delta_{h,t}$,
$\Delta_{s,t}$, and $\Delta_{k,t}$ for $t=1,...,10$ under four different cases in (\ref{error_cases}), where the data are generated from the standard GARCH model in (\ref{simgarch}) with $\eta_t\sim N(0, 1)$. In each boxplot, the lines from top to bottom represent the maximum, first quartile, median, third quartile, and maximum of the data, and the outliers are plotted individually using the `o' marker symbol.}
\label{fig-simulation_garch_normal}
\end{figure}

\begin{figure}[!h]
\centering
\includegraphics[width=150mm,height=120mm]{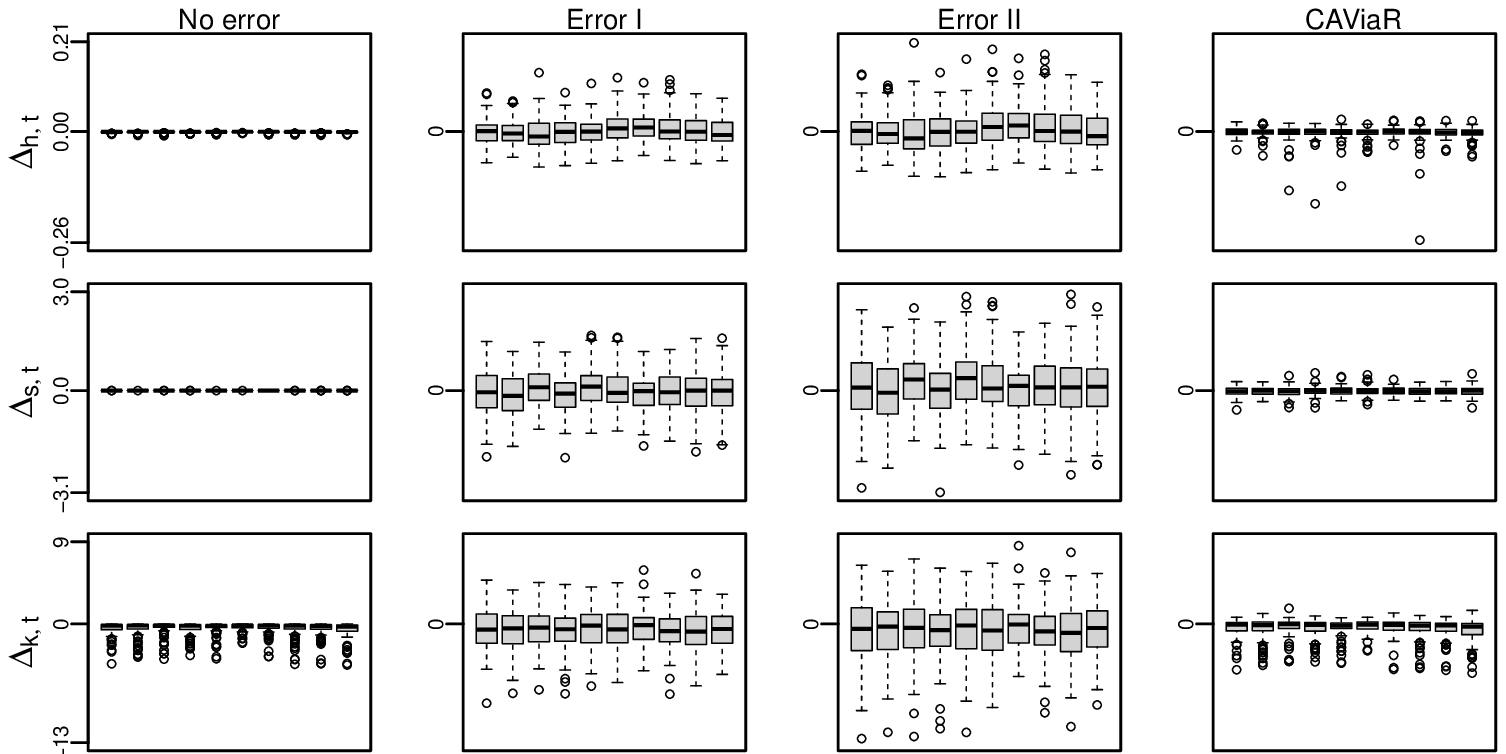}
\caption{
As for Fig \ref{fig-simulation_garch_normal}, where the data are generated from the standard GARCH model in (\ref{simgarch}) with $\eta_t\sim ST_{\nu_t}$.
}
\label{fig-simulation_garch_t}
\end{figure}

Using the values of $\{\widehat{Q}_{t}(\alpha_i)\}$ generated by (\ref{error_cases}), we compute three QCMs $\widehat{h}_t$, $\widehat{s}_t$,
and $\widehat{k}_t$ for each replication, and then we measure their precision at each $t$ by considering
\begin{equation} \label{diff}
\begin{aligned}
\Delta_{h,t}=\widehat{h}_t-h_t,\,\,\,\Delta_{s,t}=\widehat{s}_t-s_t,\,\,\,\Delta_{k,t}=\widehat{k}_t-k_t,
\end{aligned}
\end{equation}
where $h_t$, $s_t$, and $k_t$ are the true values of three CMs of $y_t$.
Based on the results of 100 replications, Figs \ref{fig-simulation_garch_normal} and \ref{fig-simulation_garch_t} exhibit the boxplots of $\Delta_{h,t}$,
$\Delta_{s,t}$, and $\Delta_{k,t}$
for $t=1,...,10$ under four different cases in (\ref{error_cases}), with $\eta_t\sim N(0, 1)$ and $ST_{\nu_t}$, respectively.
 The corresponding boxplots for
$t\geq 11$ are similar and they thus are not reported here to make the figure more visible.
From these two figures, our findings are as follows:
\begin{enumerate}
\item When there has no approximation errors as in Case 1, $\widehat{h}_t$ and $\widehat{s}_t$ are very accurate estimators of $h_t$ and $s_t$, regardless of the negligibility of the expansion errors. However, $\widehat{k}_t$ exhibits a large dispersion when
    the expansion errors are non-negligible in the case of
    $\eta_t\sim ST_{\nu_t}$, even though it has a very small dispersion when
    the expansion errors are negligible in the case of $\eta_t\sim N(0, 1)$.
    This indicates that the expansion errors typically have a minor effect on $\widehat{h}_t$ and $\widehat{s}_t$, while their impact on $\widehat{k}_t$ could be more significant.


\item When there has approximation errors with zero means (or nonzero means) as in Case 2 (or Case 3), the median lines in the boxplots of $\Delta_{h,t}$, $\Delta_{s,t}$, and $\Delta_{k,t}$ are generally close to zero, irrespective of the negligibility of the expansion errors.
    These results suggest that the three QCMs are still consistent, even when $\widehat{Q}_{t}(\alpha_i)$ are biased estimators of $Q_{t}(\alpha_i)$ and the expansion errors are present. Compared with the results in the case of $\eta_t\sim N(0, 1)$, it is expected to see that the dispersion of all QCMs becomes larger in the case of $\eta_t\sim ST_{\nu_t}$, and this phenomenon is more evident for $\widehat{k}_t$.



\item When $\widehat{Q}_{t}(\alpha_i)$ are estimated by using the CAViaR method as in Case 4, the boxplots of $\Delta_{h,t}$, $\Delta_{s,t}$, and $\Delta_{k,t}$
show that the three QCMs are consistent across all considered situations.
    Surprisingly, when $\eta_t\sim ST_{\nu_t}$,
     the performance of $\widehat{h}_t$, $\widehat{s}_t$, and $\widehat{k}_t$ in Case 4 is even better than that in Case 2 or 3. This observation is probably because the approximation errors and expansion errors are canceled out in Case 4, leading to smaller gross errors in regression model (\ref{reg3}) and so more accurate QCMs consequently.
     Therefore, our QCM method can effectively accommodate the co-existence of approximation errors and expansion errors,
     which is frequently encountered in real data analysis.

\end{enumerate}

%
%

\subsection{Simulation on the ARMA--MN--GARCH model}
Let $MN(\lambda_1,\lambda_2,\tau_1,\tau_2,\sigma_1^2,\sigma_2^2)$ denote a mixed normal (MN) distribution, the density of which
is a mixture of two normal densities of $N(\tau_1,\sigma_{1}^2)$ and $N(\tau_2,\sigma_{2}^2)$ with the weighting probabilities $\lambda_1$ and $\lambda_2$, respectively, where $\lambda_i\in(0,1)$, $i=1,2$, and $\lambda_1+\lambda_2=1$.
To examine the performance of QCMs in the presence of conditional mean specification and time-varying CMs,
we generate 100 replications of sample size $T=1000$ from the following ARMA--MN--GARCH model (\citealp{Haas:2004})
	\begin{equation}\label{simmodel}
		y_t=a_0+a_1y_{t-1}+\epsilon_t+b_1\epsilon_{t-1},
	\end{equation}
	where $\epsilon_t\sim MN(\lambda_1,\lambda_2,\tau_1,\tau_2,\sigma_{1,t}^2,\sigma_{2,t}^2)$ with
	$\tau_2=-(\lambda_1/\lambda_2)\tau_1$, $\sigma_{1,t}^2=c_{10}+c_{11}\epsilon_{t-1}^2+c_{12}\sigma_{1,t-1}^2$, and
$\sigma_{2,t}^2=c_{20}+c_{21}\epsilon_{t-1}^2+c_{22}\sigma_{2,t-1}^2$, and the parameters are chosen as
$\lambda_1=0.2$, $\tau_1=0.4$, $a_0 = 0.5$, $a_1 = 0.4$, $b_1 = -0.3$, $c_{10}=0.1$, $c_{20}=0.3$,
$c_{11}=0.05$, $c_{21}=0.1$, $c_{12}=0.85$, and $c_{22}=0.8$. For each replication, we compute the true values of CMs and conditional quantiles of $y_t$ in model (\ref{simmodel}) as follows:
	\begin{equation*} 
		\begin{aligned}
			\mu_t&=a_0+a_1y_{t-1}+b_1\epsilon_{t-1},\\
			h_t&=\lambda_1(\tau_1^2 + \sigma_{1,t}^2)+\lambda_2(\tau_2^2 + \sigma_{2,t}^2)-(\lambda_1\tau_1+\lambda_2\tau_2)^2,\\
			s_t&= \frac{\lambda_1(\tau_1^3 + 3 \tau_1\sigma_{1,t}^2)+\lambda_2(\tau_2^3 + 3 \tau_2\sigma_{2,t}^2)}{[\lambda_1 (\tau_1^2 + \sigma_{1,t}^2) +\lambda_2 (\tau_2^2 + \sigma_{2,t}^2)]^{3/2}},\\
			k_t&=\frac{\lambda_1(\tau_1^4 + 6 \tau_1^2\sigma_{1,t}^2 + 3\sigma_{1,t}^4) +\lambda_2(\tau_2^4 + 6\tau_2^2\sigma_{2,t}^2 + 3\sigma_{2,t}^4)}{
				[\lambda_1(\tau_1^2 + \sigma_{1,t}^2) +\lambda_2(\tau_2^2+\sigma_{2,t}^2)]^2},
		\end{aligned}
	\end{equation*}
and
\begin{equation*} 
	Q_t(\alpha)=\mu_t+Q_t^{\epsilon}(\alpha),
\end{equation*}
where $Q_t^{\epsilon}(\alpha)$ satisfies $\lambda_1\Phi(Q_t^{\epsilon}(\alpha);\tau_1,\sigma_{1,t}^2)+\lambda_2\Phi(Q_t^{\epsilon}(\alpha);\tau_2,\sigma_{2,t}^2)=\alpha$,
and $\Phi(x;\tau,\sigma^2)$ represents the normal distribution function with mean $\tau$ and variance $\sigma^2$.

\begin{figure}[!h]
\centering
\includegraphics[width=150mm,height=120mm]{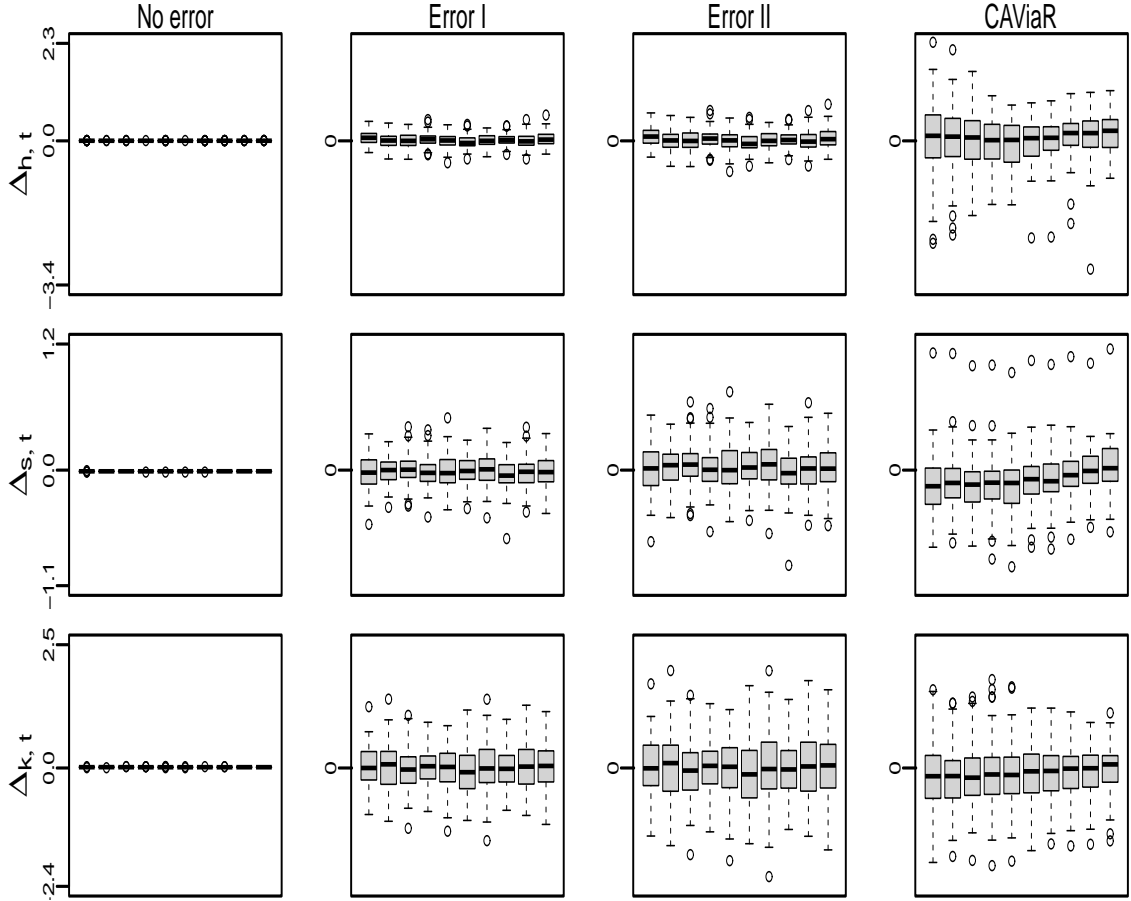}
\caption{As for Fig \ref{fig-simulation_garch_normal}, where the data are generated from the ARMA-MN-GARCH model in (\ref{simmodel}).}
\label{fig-simulation_garch_mn}
\end{figure}

Based on the results of 100 replications, Fig \ref{fig-simulation_garch_mn} exhibits the boxplots of $\Delta_{h,t}$,
$\Delta_{s,t}$, and $\Delta_{k,t}$
for $t=1,...,10$ under four different cases in (\ref{error_cases}). From this figure, we can reach the similar conclusions as in Fig \ref{fig-simulation_garch_normal}.
Hence, although $y_t$ has a heavier tail than normal under model (\ref{simmodel}), the presence of conditional mean specification does not affect the performance of QCMs.



\section{A Real Application}\label{sec_6}

In our empirical work, we consider the log-return (in percentage) series of three exchange rates, including the AUD to USD (AUD/USD), NZD to USD (NZD/USD), and CAD to USD (CAD/USD).
We denote each log-return series by $\{y_1,...,y_T\}$, which are computed from January 1, 2009 to April 20, 2023. See Table \ref{tab-summary} for some basic descriptive statistics of all three log-return series.
Below, we compute the three QCMs of each log-return series, and then use these QCMs to
study the ``News impact curve'' (NIC).

	\begin{table}[!h]
		\centering
		\caption{Descriptive statistics for three return series.}
		\label{tab-summary}
		\setlength{\tabcolsep}{11mm}{
			\begin{threeparttable}
				\begin{tabular}{lccc}
					\hline
					& AUD/USD & NZD/USD &CAD/USD\\
					\hline
					Sample size&3730&3730&3730\\
					Sample mean&$-$0.0012&0.0015&$-$0.0027\\
    					Sample variance&0.4374&0.4612&0.2056\\
   					Sample skewness&$-$0.4160&$-$0.4485&$-$0.0533\\
    					Sample kurtosis&7.0748&7.1818&5.3841\\
					\hline
				\end{tabular}
			\end{threeparttable}
		}
	\end{table}


\subsection{The three QCMs of return series}\label{sec_6_1}

Following the steps in Procedure 4.1, we compute the three QCMs $\widehat{h}_t$, $\widehat{s}_t$, and $\widehat{k}_t$ of each log-return series,
and report their basic descriptive statistics in Table \ref{tab-qcm}, where the constraint (\ref{equiv_cond_qcr}) holds for all of the computed QCMs. From Tables \ref{tab-summary} and \ref{tab-qcm}, we find that for each return series,
the mean of $\widehat{h}_t$ (or $\widehat{s}_t$) is close to the corresponding sample variance (or skewness), whereas
the mean of $\widehat{k}_t$ is much smaller than the corresponding sample kurtosis.
These findings are expected, since extreme returns can affect the sample kurtosis
for a prolonged period of time, but their impact on $\widehat{k}_t$ decays exponentially over time.

\begin{table}[!h]
  \centering
  \caption{Descriptive statistics for the three QCMs of three return series.}
  \label{tab-qcm}
  \setlength{\tabcolsep}{9.5mm}{
   \begin{threeparttable}
\begin{tabular}{ccccc}
\toprule
 & & AUD/USD & NZD/USD &CAD/USD  \\
 \midrule
\multirow{4}{*}{$\widehat{h}_t$}
 &Mean&0.5400&0.5821&0.2951\\
&Maximum& 3.0661&2.7224&1.2098\\
&Minimum&0.1380&0.1537&0.0784\\
&Ljung-Box$^\dagger$&0.0000&0.0000&0.0000\\
\midrule
\multirow{4}{*}{$\widehat{s}_t$}
&Mean&$-$0.2051&$-$0.1274&$-$0.1149\\
&Maximum&0.0463&0.1251&0.1230\\
&Minimum&$-$0.5176&$-$0.4867&$-$0.3951\\
&Ljung-Box&0.0000&0.0000&0.0000\\
\midrule
\multirow{4}{*}{$\widehat{k}_t$}
&Mean&3.7057&3.5129&3.6003\\
&Maximum&5.0147&4.7127&4.6095\\
&Minimum&2.8841&2.6247&2.8776\\
&Ljung-Box&0.0000&0.0000&0.0000\\
\bottomrule
    \end{tabular}
   \end{threeparttable}
  }
		\begin{tablenotes}
		\item $\dagger$ The results are the p-values of Ljung-Box test (\citealp{Ljung:1978}).
		\end{tablenotes}
 \end{table}

Next, we check the validity of QCMs via a similar method as in \citet{GKX:2020}.
Denote $\alpha^{h}=E(e_t^{h})$, $\alpha^{s}=E(e_t^{s})$, and $\alpha^{k}=E(e_t^{k})$, where $e_t^h=(y_t-\mu_t)^2-h_t$, $e_t^s=[(y_t-\mu_t)/\sqrt{h_t}]^3-s_t$, and $e_t^k=[(y_t-\mu_t)/\sqrt{h_t}]^4-k_t$.
Based on the estimates $\widehat{e}_t^h=(y_t-\widehat{\mu}_t)^2-\widehat{h}_t$, $\widehat{e}_t^s=[(y_t-\widehat{\mu}_t)/\sqrt{\widehat{h}_t}]^3-\widehat{s}_t$, and $\widehat{e}_t^k=[(y_t-\widehat{\mu}_t)/\sqrt{\widehat{h}_t}]^4-\widehat{k}_t$, we utilize the Student's t tests $\mathbb{T}^{h}$, $\mathbb{T}^{s}$, and $\mathbb{T}^{k}$ to test the null hypotheses $\mathbb{H}^{h}$: $a^{h}=0$, $\mathbb{H}^{s}$: $a^{s}=0$, and $\mathbb{H}^{k}$: $a^{k}=0$, respectively. Here, $\widehat{\mu}_t$ is the estimate of conditional mean, and it is computed based on the mean specifications in Section \ref{sec_6_2} below. If $\mathbb{H}^{h}$ is not rejected by $\mathbb{T}^{h}$ at the significance level $\alpha^{*}$, then it is reasonable to conclude that $\widehat{h}_t$ is valid. Similarly, the validity of
$\widehat{s}_t$ and $\widehat{k}_t$ can be examined by using $\mathbb{T}^{s}$ and $\mathbb{T}^{k}$. Table \ref{tab-student} reports the p-values of $\mathbb{T}^{h}$, $\mathbb{T}^{s}$, and $\mathbb{T}^{k}$ for all three exchange rates, and the results imply that all QCMs are valid at the significance level $5\%$.

\begin{table}[!h]
		\centering
		\caption{The p-values of $\mathbb{T}^{h}$, $\mathbb{T}^{s}$, and $\mathbb{T}^{k}$  for checking the validity of QCMs.}
		\label{tab-student}
		\setlength{\tabcolsep}{14mm}{
			\begin{threeparttable}
				\begin{tabular}{lccc}
					\hline
					&  AUD/USD & NZD/USD &CAD/USD\\
					\hline
$\mathbb{T}^{h}$	&0.4543	&0.6207 	&0.1306\\
$\mathbb{T}^{s}$	&0.3250	&0.2634	&0.3901\\
$\mathbb{T}^{k}$	&0.8321	&0.3900	&0.6425\\
					\hline
				\end{tabular}
			\end{threeparttable}
		}
	\end{table}

\begin{figure}[!h]
\centering
\includegraphics[width=170mm,height=70mm]{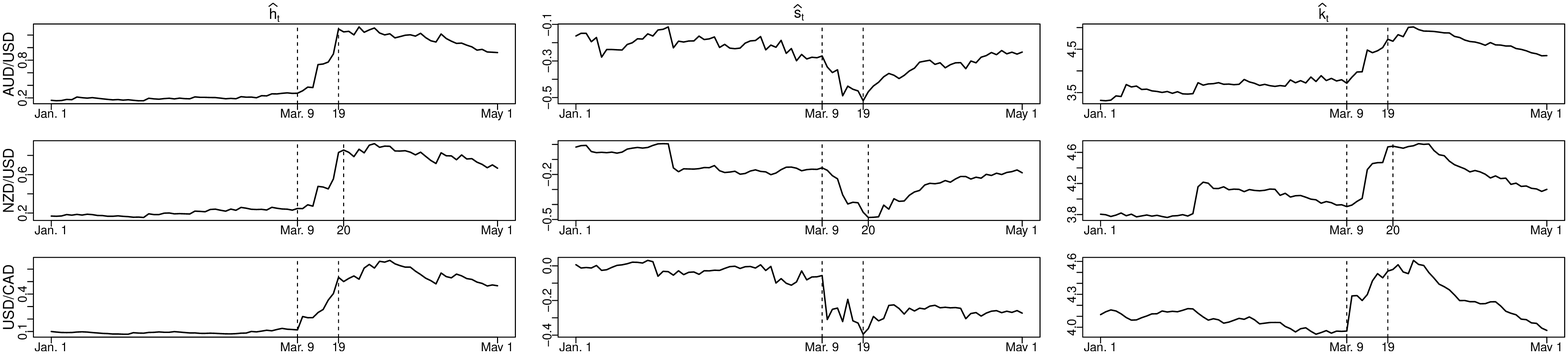}
\caption{The plots of $\widehat{h}_t$, $\widehat{s}_t$, and $\widehat{k}_t$ of three return series from January 1, 2020 to May 1, 2020.}
\label{fig-qcm}
\end{figure}

After checking the validity of the QCMs, we further plot the QCMs of all three return series
during a sub-period from January 1, 2020 to May 1, 2020 in Fig \ref{fig-qcm}. This sub-period deserves a detailed study, since it covers the 2020 stock market crash caused by the COVID-19 pandemic. For ease of visualization, the plots of all computed QCMs for the entire examined time period are not given here but available upon request. From Fig \ref{fig-qcm}, we can have the following interesting findings:

\begin{enumerate}
\item Starting from March 9, there is a rapid rising trend for both $\widehat{h}_t$ and $\widehat{k}_t$ while an apparent decline trend for $\widehat{s}_t$ in all of examined exchange rates. These observed trends demonstrate that the volatility risk kept rising sharply in currency markets, and synchronically, the tail risk to have extremely negative returns kept increasing substantially to make things even worse. The aforementioned phenomenon is not surprising, given the global impact of the COVID-19 pandemic outbreak in early 2020, which led to the depreciation of exchange rates across almost all countries afterwards.

 \item After March 19 or March 20, all of $\widehat{h}_t$, $\widehat{s}_t$, and $\widehat{k}_t$ exhibited opposite trends. This is very likely because the Federal Reserve and many central banks announced their financial rescue plans on March 17. Hence, the trend reversal sheds light on the effectiveness of the issued financial plans in rescuing stock markets.


\end{enumerate}

Overall, we find that the COVID-19 pandemic has a perilous impact on the examined three exchange rates, and the financial rescue plans are effective in reducing the values of conditional variance and kurtosis, while increasing the values of conditional skewness.

\subsection{The study on NICs}\label{sec_6_2}

 The NIC initiated by \citet{Engle:1993} aims to study how past shocks (or news) $\{\epsilon_{i}\}_{i\leq t-1}$ affect the present conditional variance $h_t$ by assuming
		\begin{equation}
		h_t = \theta_hh_{t-1}+g_h(\epsilon_{t-1}),
		\label{arch}
	\end{equation}
	where $\epsilon_{t}\equiv y_{t}-\mu_t$ is the collective shock at $t$, $\theta_h\in(0,1)$ is an unknown parameter to measure the persistence of $h_t$, and $g_h(\cdot)$ is the NIC function for $h_t$ that has a specific parametric form. For example, researchers commonly assume that
\begin{align}
g_h(x)&=\vartheta_{h,0}+\vartheta_{h,1}x^2, \label{6_model_1}\\
g_h(x)&=\vartheta_{h,0}+\vartheta_{h,1}x^2+\vartheta_{h,2}x^2\mbox{I}(x<0),\label{6_model_2}
\end{align}
where the specifications of $g_h(\cdot)$ in (\ref{6_model_1}) and (\ref{6_model_2}) lead to the standard GARCH
(\citealp{Bollerslev:1986}) and GJR-GARCH (\citealp{Glosten:1993}) models for $h_t$, respectively.
Similar to the NIC for $h_t$ in (\ref{arch}), we can follow the ideas of
\citet{Harvey:1999} and \citet{Leon:2005} to consider the NICs for $s_t$ and $k_t$:
\begin{align}
			s_t = \theta_ss_{t-1}+g_s(\varrho_{t-1}), \label{s-nic}\\
			k_t = \theta_kk_{t-1}+g_k(\varrho_{t-1}), \label{k-nic}
\end{align}
where $\varrho_{t}\equiv \epsilon_{t}/\sqrt{h_{t}}$ is the re-scaled collective shock at $t$,
$\theta_s\in (-1,1)$ and $\theta_k\in(0,1)$ are two unknown parameters to measure the persistence of $s_t$ and $k_t$, respectively, and
$g_s(\cdot)$ and $g_k(\cdot)$ are the NIC functions for $s_t$ and $k_t$, respectively.  As for $g_h(\cdot)$, $g_s(\cdot)$ and $g_k(\cdot)$  are often assumed to have certain parametric forms, such as
\begin{align}
g_s(x)&=\vartheta_{s,0}+\vartheta_{s,1}x^3, \label{6_model_3}\\
g_k(x)&=\vartheta_{k,0}+\vartheta_{k,1}x^4; \label{6_model_4}
\end{align}
see, for example, \citet{Harvey:1999} and \citet{Leon:2005}.
Since $h_t$, $s_t$, $k_t$, $\epsilon_{t}$, and $\varrho_t$ are generally unobserved, all of the unknown parameters in (\ref{arch}) and (\ref{s-nic})--(\ref{k-nic}) have to be estimated by specifying some parametric models on $y_{t}$ that account for the conditional variance, skewness, and kurtosis simultaneously.

However, so far the parametric forms of $g_h(\cdot)$, $g_s(\cdot)$, and $g_k(\cdot)$ are chosen in an ad-hoc rather than a data-driven manner.
Intuitively, if $g_h(\cdot)$, $g_s(\cdot)$, and $g_k(\cdot)$ can be estimated non-parametrically, we are able to get some useful information on their parametric forms. Motivated by this idea, we replace $h_t$, $s_t$, $k_t$, $\epsilon_{t-1}$, and $\varrho_{t-1}$ in (\ref{arch}) and
(\ref{s-nic})--(\ref{k-nic}) with $\widehat{h}_t$, $\widehat{s}_t$, $\widehat{k}_t$, $\widehat{\epsilon}_{t-1}$, and $\widehat{\varrho}_{t-1}$, respectively, where
$\widehat{\epsilon}_{t}=y_{t}-\widehat{\mu}_t$ and $\widehat{\varrho}_{t}=\widehat{\epsilon}_{t}/\sqrt{\widehat{h}_t}$ with $\widehat{\mu}_t$ being an estimator of $\mu_t$.
After this replacement, we can get the following models:
	\begin{align}
		\widehat{h}_t& = \theta_h\widehat{h}_{t-1} +g_h(\widehat{\epsilon}_{t-1})+\varsigma_{h,t}, \label{sqh}\\
        \widehat{s}_t& = \theta_s\widehat{s}_{t-1} +g_s(\widehat{\varrho}_{t-1})+\varsigma_{s,t}, \label{sqs}\\
        \widehat{k}_t& = \theta_k\widehat{k}_{t-1} +g_k(\widehat{\varrho}_{t-1})+\varsigma_{k,t}, \label{sqk}
	\end{align}
where $\varsigma_{h,t}$, $\varsigma_{s,t}$, and $\varsigma_{k,t}$ are model errors caused by the replacement.
Since the QCMs are most likely consistent estimators of CMs,  all model errors are expected to have desirable properties for valid model estimations when the parametric form of $\mu_t$ is correctly specified.

To get the correct specification of $\mu_t$, we apply two Cram\'{e}r-von Mises tests $D_{T,I}^2$ and $D_{T,C}^2$ in \citet{Escanciano:2006} to check whether the assumed form of $\mu_t$ is correctly specified, where the p-values of $D_{T,I}^2$ and $D_{T,C}^2$ are computed via the bootstrap method in \citet{Escanciano:2006}.
Since the strong autocorrelations are detected for the three return series,
we adopt an order $p$ threshold autoregressive (TAR($p$)) model (\citealp{Tong:1978}) with threshold variable being zero and delay variable being one to fit these three return series. After dismissing insignificant parameters, the AUD/USD, NZD/USD, and CAD/USD exchange rates are fitted by the TAR(5), TAR(9), and TAR(6) models, respectively. The p-values of $D_{T,I}^2$ and $D_{T,C}^2$ in
Table \ref{tab-cvmtest} indicate that these TAR models are correct specifications for the three return series at the significance level 5\%.

\begin{table}[!h]
\centering
\caption{The p-values of $D_{T,I}^2$ and $D_{T,C}^2$ for checking the conditional mean specification.}
\label{tab-cvmtest}
\begin{threeparttable}
\setlength{\tabcolsep}{14mm}{
\begin{tabular}{cccc}
\toprule
 & AUD/USD& NZD/USD & CAD/USD\\
 \midrule
$D_{T,I}^2$&0.7700&0.5400&0.4200\\
$D_{T,C}^2$&0.7200&0.4500&0.4100\\
\bottomrule
\end{tabular}}
\end{threeparttable}
\end{table}

After estimating the chosen specifications of $\mu_t$ above by the least squares method, we are able to obtain $\widehat{\epsilon}_{t-1}$ and $\widehat{\varrho}_{t-1}$.
Define $K_b(\cdot)=K(\cdot/b)/b$, where $K(\cdot)$ is Gaussian kernel function and $b$ is the bandwidth. Then, based on the sample sequence $\{(\widehat{h}_t, \widehat{h}_{t-1}, \widehat{\epsilon}_{t-1})\}_{t=2}^{T}$, we use the method in \cite{Robinson:1988} to estimate $\theta_h$ by
$$\widehat{\theta}_h=\Big\{\sum_{t=2}^{T}[\widehat{h}_{t-1}-\phi_{1}(\widehat{\epsilon}_{t-1})]^2\Big\}^{-1}
\Big\{\sum_{t=2}^{T}[\widehat{h}_{t-1}-\phi_{1}(\widehat{\epsilon}_{t-1})][\widehat{h}_{t}-\phi_{2}(\widehat{\epsilon}_{t-1})]\Big\},$$
where
$$\phi_{1}(\cdot)=\frac{\sum_{s=2}^{T}K_{b_1}(\cdot-\widehat{\epsilon}_{s-1})\widehat{h}_{s-1}}{\sum_{s=2}^{T}K_{b_1}(\cdot-\widehat{\epsilon}_{s-1})}
\mbox{ and }\phi_{2}(\cdot)=\frac{\sum_{s=2}^{T}K_{b_2}(\cdot-\widehat{\epsilon}_{s-1})\widehat{h}_{s}}{\sum_{s=2}^{T}K_{b_2}(\cdot-\widehat{\epsilon}_{s-1})},$$
and the values of $b_1$ and $b_2$ are chosen by the conventional cross-validation method.
Next, we estimate $g_h(\cdot)$ non-parametrically by
$\widehat{g}_{h}(\cdot)=\sum_{s=2}^{T}K_{b_3}(\cdot-\widehat{\epsilon}_{s-1})R_{h,s}/\sum_{s=2}^{T}K_{b_3}(\cdot-\widehat{\epsilon}_{s-1})$, where $R_{h,t}=\widehat{h}_t-\widehat{\theta}_h\widehat{h}_{t-1}$, and the value of $b_3$ is chosen by the cross-validation method. Similarly, based on the sample sequences $\{(\widehat{s}_t, \widehat{s}_{t-1}, \widehat{\varrho}_{t-1})\}_{t=2}^{T}$ and $\{(\widehat{k}_t, \widehat{k}_{t-1}, \widehat{\varrho}_{t-1})\}_{t=2}^{T}$, we estimate $g_{s}(\cdot)$ and $g_{k}(\cdot)$  non-parametrically by $\widehat{g}_{s}(\cdot)$ and $\widehat{g}_{k}(\cdot)$,  respectively.

\begin{figure}[!htp]
	\centering
	\includegraphics[width=160mm,height=160mm]{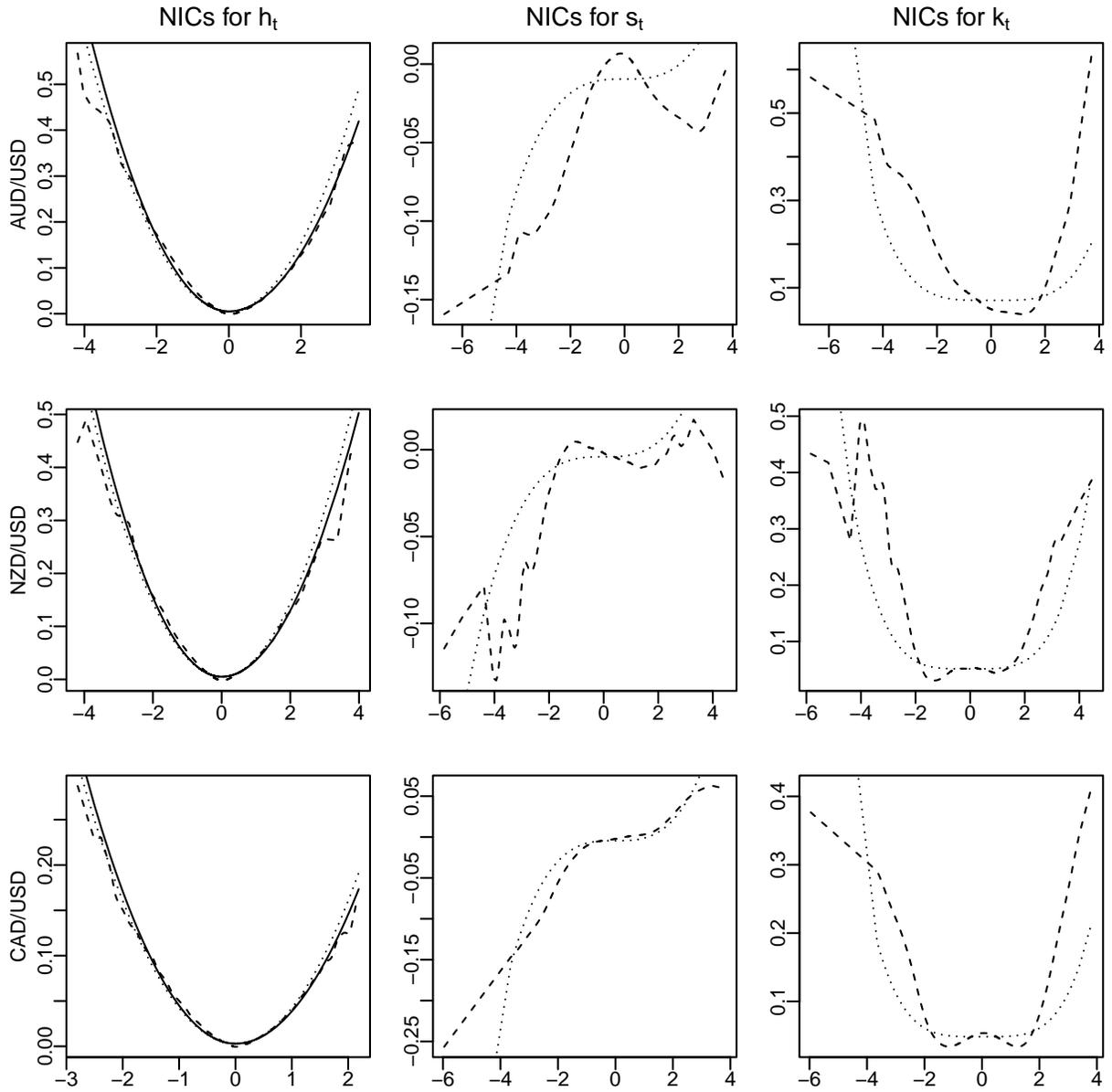}
	\caption{The plots of all fitted NICs for $h_t$, $s_t$, and $k_t$. Left panels: the non-parametric $\widehat{g}_{h}(\cdot)$ (dashed lines); the parametric $g_{h}(\cdot)$ in (\ref{6_model_1}) (dotted lines) and  (\ref{6_model_2}) (solid lines).
Middle panels: the non-parametric $\widehat{g}_{s}(\cdot)$ (dashed lines); the parametric $g_{s}(\cdot)$ in (\ref{6_model_3}) (dotted lines). Right panels: the non-parametric $\widehat{g}_{k}(\cdot)$ (dashed lines); the parametric $g_{k}(\cdot)$ in (\ref{6_model_4}) (dotted lines).
}
	\label{fig-nic}
	\end{figure}

Fig \ref{fig-nic} plots the non-parametric fitted  models $\widehat{g}_{h}(\cdot)$, $\widehat{g}_{s}(\cdot)$, and $\widehat{g}_{k}(\cdot)$ for all three return series. From this figure, we find that the form of $g_h(\cdot)$ in (\ref{6_model_1}) or (\ref{6_model_2}) matches $\widehat{g}_{h}(\cdot)$ quite well, whereas the forms of $g_{s}(\cdot)$ in (\ref{6_model_3}) and $g_{k}(\cdot)$ in (\ref{6_model_4}) exhibit a large deviation from $\widehat{g}_{s}(\cdot)$ and $\widehat{g}_{k}(\cdot)$, respectively, in all three cases. The same conclusion can be reached in view of the results of adjusted R$^2$ for all fitted models in Table \ref{tab-rsquare}.

\begin{table}[!h]
		\centering
		\caption{The values of adjusted R$^2$ for the fitted models (\ref{sqh})--(\ref{sqk}).}
		\label{tab-rsquare}
			\begin{threeparttable}
			\setlength{\tabcolsep}{12.3mm}{
\begin{tabular}{lccc}
\toprule

 & AUD/USD & NZD/USD & CAD/USD \\
\midrule

&\multicolumn{3}{c}{Panel A: Model (\ref{sqh})}  \\
$g_h(\cdot)\sim(\ref{6_model_1})$ &0.9215  &0.9270  &0.9551\\
$g_h(\cdot)\sim(\ref{6_model_2})$ &0.9363 & 0.9357  &0.9628\\
\midrule

&\multicolumn{3}{c}{Panel B: Model (\ref{sqs})}  \\
$g_s(\cdot)\sim(\ref{6_model_3})$ &  0.0920&0.1316&0.4754\\
\midrule

&\multicolumn{3}{c}{Panel C: Model (\ref{sqk})}  \\
$g_k(\cdot)\sim(\ref{6_model_4})$ &0.1416  &0.3008 &0.2702\\
\bottomrule
\end{tabular}}
		\end{threeparttable}
	\end{table}

\section{Concluding Remarks}\label{sec_7}
	This paper estimates the three CMs (with respect to variance, skewness, and kurtosis) by the corresponding QCMs, which are easily computed from the OLS estimator of a linear regression model constructed by the ECQs. The QCM method builds on the Cornish-Fisher expansion, which essentially transforms the estimation of CM to that of conditional quantile.
Such transformation brings us two attractive advantages over the parametric GARCH-type methods:
First, the QCM method bypasses the investigation of conditional mean and allows the mis-specified conditional quantile models, due to
its regression-based feature; Second, the QCM method has stable estimation results, since it does not involve any complex nonlinear constraint on the admission region of parameters in conditional quantile models.
These two advantages come with two limitations. The first limitation is that we need to do the conditional quantile estimation $n$ different times. But this seems neither a computational burden nor a theoretical obstruct. The second limitation is that when the data are more heavy-tailed, the CF expansion is unavoidable to be less accurate, leading to more non-negligible expansion error. Although this limitation
does not affect the consistency of the QCMs in general, it makes the QCMs (especially the quantiled conditional kurtosis) have a larger dispersion when the data are more heavy-tailed.

Notably, the QCM method is a supervised learning procedure without assuming any distribution on the returns. This important supervised learning feature makes the QCM method have a substantially computational advantage over the GARCH-type methods to
estimate conditional variance, skewness, and kurtosis, when the dimension of return is large.
See \cite{ZZZ:2023} for an in-depth discussion on this context and an innovative method for big portfolio selections based on the
learned conditional higher-moments from the QCM method.

Finally, we should mention that the existing parametric methods commonly only work for stationary data and their extension to deal with more complex data environments seems challenging in terms of methodology and computation.
In contrast, the QCM method could be applicable in complex data environments, as long as the ECQs are fairly provided. For example, the QCMs could adapt to the mixed categorical and continuous data or locally stationary data when the ECQs are computed by the method in \cite{LR:2008} or \cite{Zhou:2009}, respectively. In addition, the useful information from exogenous variables and conditional quantiles of other variables can be easily embedded to the QCMs through the channels of the ECQs as done in \cite{HWY:2016} and \cite{TB:2016}. Since the QCMs are computed at each fixed timepoint,
the QCM method also allows us to pay special attention to the CMs during a specific time period by employing the methods in \cite{Cai:2002} and \cite{XU:2013} to compute the ECQs. On the whole, the QCM method exhibits a much wider application scope than the parametric ones, which
so far have not offered a clear feasible manner to study the CMs under the above complex data environments.

\bibliographystyle{imsart-nameyear}

\begin{thebibliography}{49}

        \bibitem[{Andrews(1988)}]{Andrews:1988}
         Andrews, D. W. K. (1988).
         \newblock Laws of large numbers for dependent nonidentically distributed random variables.
         \newblock {\it Econometric Theory} {\bf 4}, 458--467.
		
        \bibitem[{Bali et al.(2008)}]{BMT:2008}
	    Bali, T. G., Mo, H. and Tang, Y. (2008).
        \newblock The role of autoregressive conditional skewness and kurtosis in the estimation of conditional VaR.
        \newblock {\it Journal of Banking \& Finance} {\bf 32}, 269--282.
		
		\bibitem[{Bollerslev(1986)}]{Bollerslev:1986}
		Bollerslev, T. (1986).
		\newblock Generalized autoregressive conditional heteroskedasticity.
		\newblock {\it Journal of Econometrics} {\bf 31}, 307--327.



		
		


        \bibitem[{Brooks et al.(2005)}]{BBHP:2005}
        Brooks, C., Burke, S. P., Heravi, S. and Persand, G. (2005).
        \newblock Autoregressive conditional kurtosis.
        \newblock {\it Journal of Financial Econometrics} {\bf 3}, 399--421.
		
         \bibitem[{Cai(2002)}]{Cai:2002}
         Cai, Z. (2002).
         \newblock Regression quantiles for time series.
         \newblock {\it Econometric Theory} {\bf 18}, 169--192.

		
		
		
		\bibitem[{Chunhachinda et al.(1997)}]{Chunhachinda:1997}
		Chunhachinda, P., Dandapani, K., Hamid, S. and Prakash, A. J. (1997).
		\newblock Portfolio selection and skewness: Evidence from international stock markets.
		\newblock {\it Journal of Banking \& Finance} {\bf 21}, 143--167.
		
		\bibitem[{Cornish and Fisher(1938)}]{Cornish:1938}
		Cornish, E. A. and Fisher, R. A. (1938).
		\newblock Moments and cumulants in the specification of distributions.
		\newblock {\it Revue de l'Institut international de Statistique} {\bf 5}, 307--320.
		


%
%
%
	
	    \bibitem[{Engle(1982)}]{Engle:1982}
		Engle, R. F. (1982).
		\newblock Autoregressive conditional heteroscedasticity with estimates of the variance of United Kingdom inflation.
		\newblock {\it Econometrica} {\bf 50}, 987--1007.
		
%


		\bibitem[{Engle and Manganelli(2004)}]{Engle:2004}
		Engle, R. F. and Manganelli, S. (2004).
		\newblock CAViaR: Conditional autoregressive value at risk by regression quantiles.
		\newblock {\it Journal of Business \& Economic Statistics} {\bf 22}, 367--381.
				
		\bibitem[{Engle and Ng(1993)}]{Engle:1993}
		Engle, R. F. and Ng, V. K. (1993).
		\newblock Measuring and testing the impact of news on volatility.
		\newblock {\it Journal of Finance} {\bf 48}, 1749--1778.
		
			
		\bibitem[{Escanciano(2006)}]{Escanciano:2006}
		Escanciano, J. C. (2006).
		\newblock Goodness-of-fit tests for linear and nonlinear time series models.
		\newblock {\it Journal of the American Statistical Association} {\bf 101}, 140--149.

        \bibitem[{Escanciano(2009)}]{Escanciano:2009}
		Escanciano, J. C. (2009).
        \newblock Quasi-maximum likelihood estimation of semi-strong GARCH models.
        \newblock {\it Econometric Theory} {\bf 25}, 561--570.

        \bibitem[{Escanciano and Velasco(2006)}]{EV:2006}
         Escanciano, J. C. and Velasco, C. (2006).
         \newblock Generalized spectral tests for the martingale difference hypothesis.
         \newblock {\it Journal of Econometrics} {\bf 134}, 151--185.

		
				
		

        \bibitem[{Fan and Yao(2003)}]{FY:2003}
        Fan, J. and Yao, Q. (2003).
        \newblock {\it Nonlinear Time Series: Nonparametric and Parametric Methods}.
        \newblock Springer, New York.
		

        \bibitem[{Francq and Sucarrat(2022)}]{FS:2022}
        Francq, C. and Sucarrat, G. (2022).
        \newblock Volatility estimation when the zero-process is nonstationary.
        \newblock {\it Journal of Business \& Economic Statistics} {\bf 41}, 53--66.
		
	    \bibitem[{Francq and Thieu(2019)}]{FT:2019}
         Francq, C. and Thieu, L. Q. (2019).
         \newblock QML inference for volatility models with covariates.
         \newblock {\it Econometric Theory} {\bf 35}, 37--72.

			
		\bibitem[{Francq and Zako\"{i}an(2019)}]{Francq:2019}
		Francq, C. and Zako\"{i}an, J. M. (2019).
		\newblock {\it GARCH Models: Structure, Statistical Inference and Financial Applications}.
		\newblock John Wiley \& Sons.
		
		

	

		
		\bibitem[{Glosten et al.(1993)}]{Glosten:1993}
		Glosten, L. R., Jagannathan, R. and Runkle, D. E. (1993).
		\newblock On the relation between the expected value and the volatility of the nominal excess return on stocks.
		\newblock {\it Journal of Finance} {\bf 48}, 1779--1801.

       \bibitem[{Grigoletto and Lisi(2009)}]{GL:2009}
         Grigoletto, M. and Lisi, F. (2009).
         \newblock Looking for skewness in financial time series.
         \newblock {\it Econometrics Journal} {\bf 12}, 310--323.
	
	\bibitem[{Gu et al.(2020)}]{GKX:2020}
         Gu, S., Kelly, B. and Xiu, D. (2020).
         \newblock Empirical asset pricing via machine learning.
         \newblock {\it Review of Financial Studies} {\bf 33}, 2223--2273.
	
	
	
		
		
		\bibitem[{Haas et al.(2004)}]{Haas:2004}
		Haas, M., Mittnik, S. and Paolella, M. S. (2004).
		\newblock Mixed normal conditional heteroskedasticity.
		\newblock {\it Journal of Financial Econometrics} {\bf 2}, 211--250.
		
		
		\bibitem[{Hansen(1994)}]{Hansen:1994}
		Hansen, B. E. (1994).
		\newblock Autoregressive conditional density estimation.
		\newblock {\it International Economic Review} {\bf 35}, 705--730.



         \bibitem[{H\"{a}rdle et al.(2016)}]{HWY:2016}
         H\"{a}rdle, W. K., Wang, W. and Yu, L. (2016).
         \newblock TENET: Tail-Event driven NETwork risk.
         \newblock {\it Journal of Econometrics} {\bf 192}, 499--513.
		
		
		
		
		\bibitem[{Harvey and Siddique(1999)}]{Harvey:1999}
		Harvey, C. R. and Siddique, A. (1999).
		\newblock Autoregressive conditional skewness.
		\newblock {\it Journal of Financial and Quantitative Analysis} {\bf 34}, 465--487.
		
		\bibitem[{Harvey and Siddique(2000)}]{Harvey:2000}
		Harvey, C. R. and Siddique, A. (2000).
		\newblock Conditional skewness in asset pricing tests.
		\newblock {\it Journal of Finance} {\bf 55}, 1263--1295.
		
		
	
		
		\bibitem[{Jondeau and Rockinger(2003)}]{Jondeau:2003}
		Jondeau, E. and Rockinger, M. (2003).
		\newblock Conditional volatility, skewness, and kurtosis: existence, persistence, and comovements.
		\newblock {\it Journal of Economic Dynamics and Control} {\bf 27}, 1699--1737.
		

        \bibitem[{Jondeau et al.(2019)}]{JZZ:2019}
        Jondeau, E., Zhang, Q. and Zhu, X. (2019).
        \newblock Average skewness matters.
        \newblock {\it Journal of Financial Economics} {\bf 134}, 29--47.

		\bibitem[{Koenker and Bassett(1978)}]{Koenker:1978}
		Koenker, R. and Bassett, G. (1978).
       	\newblock Regression quantiles.
        \newblock {\it Econometrica} {\bf 46}, 33--50.


        \bibitem[{Koenker et al.(2017)}]{KCHP:2017}
        Koenker, R., Chernozhukov, V., He, X. and Peng, L. (2017).
        \newblock {\it Handbook of Quantile Regression}.
        \newblock Chapman \& Hall/CRC.
		



        \bibitem[{Kuester et al.(2006)}]{KMP:2006}
        Kuester, K., Mittnik, S. and Paolella, M. S. (2006).
        \newblock Value-at-risk prediction: A comparison of alternative strategies.
        \newblock {\it Journal of Financial Econometrics} {\bf 4}, 53--89.
		


\bibitem[{Lee and Lin(1992)}]{LeeLin:1992}
Lee, Y. S. and Lin, T. K. (1992).
\newblock Algorithm AS 269: High order Cornish-Fisher expansion.
\newblock {\it Journal of the Royal Statistical Society: Series C} {\bf 41}, 233--240.
		

        \bibitem[{Le\'{o}n and \={N}\'{i}guez(2020)}]{LN:2020}
        Le\'{o}n, \'{A}. and \={N}\'{i}guez, T. M. (2020).
        \newblock Modeling asset returns under time-varying semi-nonparametric distributions.
        \newblock {\it Journal of Banking \& Finance} {\bf 118}, 105870.
		
		\bibitem[{Le\'{o}n et al.(2005)}]{Leon:2005}
		Le\'{o}n, \'{A}., Rubio, G. and Serna, G. (2005).
		\newblock Autoregresive conditional volatility, skewness and kurtosis.
		\newblock {\it Quarterly Review of Economics and Finance} {\bf 45}, 599--618.


        \bibitem[{Li and Racine(2008)}]{LR:2008}
         Li, Q. and Racine, J. S. (2008).
         \newblock Nonparametric estimation of conditional CDF and quantile functions with mixed categorical and continuous data.
         \newblock {\it Journal of Business \& Economic Statistics} {\bf 26}, 423--434.

		
		\bibitem[{Ljung and Box(1978)}]{Ljung:1978}
		Ljung, G. M. and Box, G. E. (1978).
		\newblock On a measure of lack of fit in time series models.
		\newblock {\it Biometrika} {\bf 65}, 297--303.




				
		 \bibitem[{McNeil and Frey(2000)}]{MF:2000}
         McNeil, A. J. and Frey, R. (2000).
         \newblock Estimation of tail-related risk measures for heteroscedastic financial time series: an extreme value approach.
         \newblock {\it Journal of Empirical Finance} {\bf 7}, 271--300.
		

		
		
		



		
		
	
		
	
		
		
		\bibitem[{Robinson(1988)}]{Robinson:1988}
		Robinson, P. M. (1988).
		\newblock Root-N-consistent semiparametric regression.
		\newblock {\it Econometrica} {\bf 56}, 931--954.
		

       \bibitem[{Rubinstein(1973)}]{Rubinstein:1973}
        Rubinstein, M. E. (1973).
        \newblock A comparative statics analysis of risk premiums.
        \newblock {\it Journal of Business} {\bf 46}, 605--615.


        \bibitem[{Samuelson(1970)}]{Samuelson:1970}
		Samuelson, P. A. (1970).
        \newblock The fundamental approximation theorem of portfolio analysis in terms of means, variances and higher moments.
        \newblock {\it Review of Economic Studies} {\bf 37}, 537--542.
		

		

        \bibitem[{Sucarrat and Gr{\o}nneberg(2022)}]{SG:2022}
		Sucarrat, G. and Gr{\o}nneberg, S. (2022).
        \newblock Risk estimation with a time-varying probability of zero returns.
        \newblock {\it Journal of Financial Econometrics} {\bf 20}, 278--309.




        \bibitem[{Tobias and Brunnermeier(2016)}]{TB:2016}
        Tobias, A. and Brunnermeier, M. K. (2016).
        \newblock CoVaR.
        \newblock {\it American Economic Review} {\bf 106}, 1705--1741.
		
		\bibitem[{Tong(1978)}]{Tong:1978}
		Tong, H. (1978).
		\newblock On a threshold model.
		\newblock {\it Pattern recognition and signal processing}. (ed. C.H.Chen). Sijthoff and Noordhoff, Amsterdam.
		

        \bibitem[{Tsay(2005)}]{Tsay:2005}
        Tsay, R. S. (2005).
        \newblock {\it Analysis of Financial Time Series}.
        \newblock John Wiley \& Sons.

		



		

		\bibitem[{White(2001)}]{White:2001}
		White, H. (2001).
		\newblock {\it Asymptotic Theory for Econometricians, Revised edition}.
		\newblock San Diego: Academic Press.
		

        \bibitem[{Widder(1946)}]{Widder:1946}
        Widder, D. V. (1946).
        \newblock {\it The Laplace Transform}.
        \newblock Princeton University Press, Princeton, NJ.

       \bibitem[{Xiao and Koenker(2009)}]{XK:2009}
        Xiao, Z. and Koenker, R. (2009).
        \newblock Conditional quantile estimation for generalized autoregressive conditional heteroscedasticity models.
        \newblock {\it Journal of the American Statistical Association} {\bf 104}, 1696--1712.



        \bibitem[{Xu(2013)}]{XU:2013}
        Xu, K. L. (2013).
        \newblock Nonparametric inference for conditional quantiles of time series.
        \newblock {\it Econometric Theory} {\bf 29}, 673--698.



        \bibitem[{Zheng at al.(2018)}]{Zheng:2018}
        Zheng, Y., Zhu, Q., Li, G. and Xiao, Z. (2018).
        \newblock Hybrid quantile regression estimation for time series models with conditional heteroscedasticity.
        \newblock {\it Journal of the Royal Statistical Society: Series B} {\bf 80}, 975--993.
		
		\bibitem[{Zhou and Wu(2009)}]{Zhou:2009}
		Zhou, Z. and Wu, W. B. (2009).
		\newblock Local linear quantile estimation for nonstationary time series.
		\newblock {\it Annals of Statistics} {\bf 37}, 2696--2729.

		
		\bibitem[{Zhu et al.(2023)}]{ZZZ:2023}
        Zhu, Z., Zhang, N. and Zhu, K. (2023).
        \newblock Big portfolio selection by graph-based conditional moments method.
        \newblock Working paper. Available at ``https://arxiv.org/abs/2301.11697''.

\end{thebibliography}

\end{document}